\documentclass[aps,prd,preprint,superscriptaddress ,nofootinbib]{revtex4-1}
\usepackage{chay,graphicx}

\newcommand{\eps}{\epsilon}
 \newcommand{\euv}{\epsilon_{\mathrm{UV}}}
\newcommand{\eir}{\epsilon_{\mathrm{IR}}}

\begin{document}
\title{Analysis of exclusive $k_T$ jet algorithms in electron-positron annihilation}

\def\KU{Department of Physics, Korea University, Seoul 136-713, Korea} 
\def\Seoultech{Institute of Convergence Fundamental Studies and School of Liberal Arts, 
Seoul National University of Science and 
Technology, Seoul 139-743, Korea}
\author{Junegone Chay}
\email[E-mail:]{chay@korea.ac.kr}
\affiliation{\KU}
\author{Chul Kim}
\email[E-mail:]{chul@seoultech.ac.kr}
\affiliation{\Seoultech}
\author{Inchol Kim} 
\email[E-mail:]{vorfeed@korea.ac.kr}
\affiliation{\KU}
 
\begin{abstract} \vspace{0.1cm}\baselineskip 3.0 ex 
 We  study the factorization of the dijet cross section in $e^+ e^-$ annihilation using the generalized exclusive jet algorithm which includes 
the cone-type, the JADE, the $k_T$, the anti-$k_T$ and the  Cambridge/Aachen jet algorithms as special cases. In order to probe 
the characteristics of the jet algorithms in a unified way, we consider the generalized $k_T$ jet algorithm with an arbitrary weight of 
the energies, in which various types of the  $k_T$-type algorithms are included for specific values of the parameter.  We show that the jet algorithm
respects the factorization property for the parameter $\alpha <2$. The factorized jet function and the soft function are well defined and 
infrared safe for all the jet algorithms except the $k_T$  algorithm. The $k_T$ 
algorithm ($\alpha=2$) breaks the factorization since the jet and the soft functions are infrared divergent and are not defined for $\alpha=2$, 
though the dijet cross section is infrared finite. 
In the jet algorithms which enable factorization, we give a phenomenological analysis using the resummed and the fixed-order results. 
\end{abstract}

\maketitle

\baselineskip 3.0 ex 

\section{Introduction}
In high-energy scattering, strong interaction plays a significant role both in the high-energy region
and in the low-energy region. The scattering of energetic quarks and gluons with  radiative corrections
occur in the high-energy regime, in which perturbative QCD can describe the effects of the strong interaction.
On the other hand, strong interaction is also responsible for pulling the quarks and gluons out of the initial
hadrons through the parton distribution functions, and for the hadronization of the energetic scattered partons
through the fragmentation functions. These are nonperturbative aspects of the strong interaction in the low-energy regime. 
Furthermore the strong interaction affects all these processes at different energy scales at the
same time, therefore the effects of the strong interaction are entangled in a complicated way 
all through the scattering process from  beginning to  end. 

It is important to disentangle the short- and the long-distance effects of the strong interaction to have 
theoretical predictive power.  This is achieved by the factorization theorem in which the scattering cross section is
factorized into the hard, collinear and soft parts. The factorization proof in various hard scattering processes
has been a long-standing problem in QCD \cite{Collins:1987pm,Collins:1989gx}.
Recently the advent of the soft-collinear effective theory (SCET) \cite{Bauer:2000ew,Bauer:2000yr,Bauer:2001yt}
facilitates the proof in a straightforward way. SCET is formulated such that the collinear and the soft modes are
decoupled at the Lagrangian level, and the factorization of the scattering cross section into the hard, collinear
and soft parts is intuitively transparent. 

In high-energy processes, final-state hadrons are typically clustered in collimated beams, which are called jets. 
Jets are important tools to exploit information on Standard Model or beyond. There are many jet definitions depending on 
the kinematics, or the detector design,  etc. \cite{Salam:2009jx}. In this paper we will confine ourselves to the study of the various jet 
definitions used in $e^+e^-$ annihilation, but it can be extended to $ep$ scattering
or to $pp$ scattering. The jet definitions are realized by providing the jet algorithms which give rules on how to combine particles in forming a jet.
The jet algorithms include the Sterman-Weinberg algorithm \cite{Sterman:1977wj}, the cone-type 
algorithm \cite{Salam:2007xv}, the JADE algorithm \cite{Bethke:1988zc}, the $k_T$ algorithm \cite{Catani:1991hj}, 
the anti-$k_T$ algorithm \cite{Cacciari:2008gp},  the Cambridge/Aachen algorithm \cite{Dokshitzer:1997in,Wobisch:1998wt} to name a few. 

The basic issue in jet algorithms is how to combine adjacent final-state particles into a jet without ambiguity. The important ingredients of 
appropriate jet algorithms are such that they should yield infrared safety, and they can be employed conveniently both in theory and 
experiment. However,
the implementation of a convenient jet algorithm in experiment is not always convenient in theory, or vice versa.

The factorized form of the dijet cross section in $e^+e^-$ annihilation can be obtained in the framework of SCET. The cross section
is factorized into the hard, collinear and soft parts. However, the factorization does not hold for any arbitrary jet algorithm if we require
that each factorized part be infrared finite. If any factorized part contains infrared divergence for a certain jet algorithm, 
though the partial sum of the factorized parts is infrared safe, the factorization approach loses its physical meaning.

In order to approach the problem about which jet algorithms respect the factorization theorem, we devise a general jet algorithm, 
motivated by the $k_T$ jet algorithm, in which we introduce a parameter $\alpha$.  The original exclusive $k_T$ algorithm is given by
\begin{equation} \label{yij}
y_{ij} =\frac{2}{Q^2} (1-\cos \theta_{ij}) \, \mathrm{min.} (E_i^2, E_j^2) <y_c,
\end{equation}
where $Q$ is the center-of-mass energy.  For a given pair of partons $i$ and $j$, with the energy $E_i$ and $E_j$ respectively, the parameter
$y_{ij}$ is constructed with the relative angle $\theta_{ij}$. 
Physically this jet algorithm requires that the relative transverse momentum of 
every pair of final-state partons $i$ and $j$ be measured. If the smallest value of $y_{ij}$ for a given pair of partons is less than a resolution 
parameter $y_c$, they are combined into a jet. This process is repeated until all pairs have $y_{ij}>y_c$.  If the relative transverse momentum is 
replaced by the invariant mass, the jet algorithm becomes the JADE algorithm. 

Here we employ the generalized exclusive jet algorithm, which is defined as
\begin{equation}  \label{genkt}
y_{ij} =\frac{2}{Q^{\alpha}} (1-\cos \theta_{ij}) \,\mathrm{min.} (E_i^{\alpha}, E_j^{\alpha}) <y_c,
\end{equation}
where the power of the energy in Eq.~(\ref{yij}) is replaced by $\alpha$, and the resolution parameter $y_c$ remains dimensionless.
As $\alpha$ varies, the general jet algorithm includes 
all the jet algorithms mentioned above for specific values of $\alpha$.  For example, the $k_T$ algorithm is obtained for $\alpha=2$. The JADE
algorithm is similar to the case $\alpha=1$. The Cambridge/Aachen algorithm corresponds to $\alpha=0$, and the 
anti-$k_T$ algorithm\footnote{\baselineskip 3.0ex The anti-$k_T$ algorithm in Ref.~\cite{Cacciari:2008gp} is proposed for the inclusive jet algorithm. 
We refer to the exclusive jet algorithm with $\alpha=-2$ in Eq.~(\ref{genkt}) as the anti-$k_T$ algorithm.}  is given
by $\alpha=-2$. We probe all the possible values of $\alpha$ and compute 
the jet and the soft functions to see if they are well defined and free
of infrared divergence.  The phase space available for the collinear and the soft functions
differs for different values of $\alpha$, so does the structure of divergence.

The next-to-leading-order contribution of each factorized part will be computed accordingly, in which
there are at most two particles in a jet. 
We compute each contribution using the dimensional regularization with the spacetime
dimension $D= 4-2\eps$ regulating both the ultraviolet (UV) and the infrared (IR) divergences. Also the 
$\overline{\mathrm{MS}}$ scheme is employed with $4\pi \mu_{\overline{\mathrm{MS}}}^2
= \mu^2 e^{\gamma_{\mathrm{E}}}$. The separation of the UV and IR divergences is important to see whether each factorized part is IR finite.
So far, it has been proven that the IR divergence cancels in many processes. And, in fact, any physical quantity should be IR safe.
In this case, the remaining divergence is of the UV origin. Technically,
we can put all the scaleless integrals to be zero, and also put all the poles in $\eps$ as the UV poles in $\euv$. But this is possible 
only after it is verified that each factorized part is IR finite. 

We intend to verify which types of the jet algorithms guarantee the IR finiteness of each factorized part, 
and which do not, using the generalized jet algorithm. 
As it turns out, the jet algorithm for $\alpha <2$ indeed yields IR-finite jet 
and soft functions, hence can safely put all the poles as the UV poles. However, for $\alpha>2$ each jet and soft function contains
IR divergence, and we cannot naively put all the poles as the UV poles, though the jet cross section is IR finite. 
 This verification is possible only when we distinguish the UV and IR 
divergences, which we explicitly show here at next-to-leading order. The existence of the IR divergence in the factorized part, or the 
breakdown of the factorization theorem was also considered in Ref.~\cite{Hornig:2009vb} in a different context.  
 
In full QCD, only the jet cross section has been computed and the divergence is regarded as the UV divergence. 
This is justified since the cross section is 
a physical quantity free of IR divergence. However, the factorization property in high-energy scattering of the jet cross section or any other 
physical observables such as thrust is important to disentangle the strong interaction and to compute high-order corrections for better accuracy.
Then it is crucial to see if the factorization works or breaks depending on the jet algorithms. We address this problem by considering the 
IR finiteness of the jet and the soft functions in various jet algorithms and point out which jet algorithms allow the factorization theorem 
in the jet cross section.

The structure of the paper is as follows: In Section~\ref{factor}, the factorization of the dijet cross section with the jet algorithm is briefly described 
in the framework of SCET.
In Section~\ref{jetalgo},  the phase space for the naive collinear, the zero-bin and the soft contributions is analyzed 
using the generalized  jet algorithm. The structure of the phase space shows different behavior for different values of $\alpha$, and we
classify them according to the possible divergence structure.

In Section~\ref{posalp}, the collinear and soft contributions for $\alpha>0$ are computed. 
These contributions are considered for various values of $\alpha$, corresponding to the different phase spaces. We verify that the collinear 
and soft functions are infrared finite for $0 <\alpha <2$, while they contain infrared divergence for $\alpha>2$. The case $\alpha=2$, corresponding
to the original $k_T$ algorithm is not well defined in dimensional regularization though it can be inferred that each contribution contains the IR
divergence from the consideration of the phase space. 
In Section~\ref{negalp}, the collinear and soft contributions are computed for negative $\alpha$.
In this case, there should be a jet veto needed in calculating the soft function. Each contribution is IR finite for any negative $\alpha$, and the
generalized jet algorithm with $\alpha<0$ shows similar behavior compared to the cone-type and the Sterman-Weinberg jet algorithms
\cite{Chay:2015ila}. In Section~\ref{psdiv}, we discuss the structure of divergence for various values of $\alpha$. The identification of the 
UV and IR divergences can be inferred from the structure of the corresponding phase space. It is explained in detail what kind of divergence
there will be {\it a priori} by examining the structure of the phase space.
In Section~\ref{alpha2}, we discuss peculiar characteristics of the $k_T$ algorithm with $\alpha=2$. In dimensional regularization, 
it is impossible to extract the poles in the jet and the soft functions. Furthermore they contain poles in $1/(\alpha-2)$.
The behavior of the collinear and soft parts for $\alpha=2$ is explained in detail.

In Section~\ref{xsec}, all the contributions are added to compute the dijet cross section. The jet algorithm with $\alpha<2$ yields finite collinear
and soft functions. For $\alpha \ge 2$, the factorization breaks down, that is, each function contains IR divergence. However, if we add all the
contributions, the IR divergence and the singularity in $1/(\alpha-2)$ cancel and the resultant dijet cross section is IR finite.  We also present the 
resummed results at next-to-leading logarithmic accuracy for the dijet cross sections with their theoretical uncertainty, and compare with
the fixed-order results. In Section~\ref{conc}, we summarize the characteristics of the jet and the soft functions in the generalized jet 
algorithm and give conclusions. In Appendix, we collect all the ingredients for the numerical analysis.

\section{Factorization of the dijet cross section\label{factor}}
The factorization theorem of the dijet cross section has been presented in Ref.~\cite{Chay:2015ila}, and we briefly review the result instead of
rederiving it.
The factorized dijet cross section is given by
\begin{equation} \label{facjet} 
\sigma_{\mathrm{jet}} =\sigma_0  H(Q^2,\mu)    \mathcal{J}_{n,\Theta} ( \mu)   \mathcal{J}_{\overline{n},\Theta} (\mu) 
\mathcal{S}_{\Theta} (\mu).
\end{equation}
Here $Q^2$ is the invariant-mass squared of the $e^+ e^-$ system, and $\sigma_0$ is the Born cross section for a given flavor $f$ 
of the quark-antiquark pair with the electric charge $Q_f$, given by
\begin{equation}
\sigma_0 = \frac{4\pi \alpha^2 Q_f^2 N_c}{3Q^2}.
\end{equation}
$H(Q^2,\mu)$ is the hard function which is obtained from the matching of the electromagnetic current between the full QCD and 
SCET at leading order as
\begin{equation}
\label{current}
J^{\mu} = C(Q^2,\mu) \overline{\chi}_n \tilde{Y}_n^{\dagger} \gamma^{\mu} \tilde{Y}_{\overline{n}} \chi_{\overline{n}},
\end{equation}
and $H(Q^2, \mu) = |C(Q^2,\mu)|^2$. To one loop, it is given by \cite{Manohar:2003vb}
\begin{equation} \label{hard}
H(Q^2,\mu) = 1+\frac{\alpha_s C_F}{2\pi}
\Bigl (-\ln^2  \frac{\mu^2}{Q^2} -3 \ln \frac{\mu^2}{Q^2} -8
+\frac{7\pi^2}{6}\Bigr). 
\end{equation}
And $\chi_n$ is a gauge-invariant collinear quark with a collinear Wilson line $\chi_n = W_n^{\dagger} \xi_n$, and $\tilde{Y} (x)$ is the soft
Wilson line \cite{Chay:2004zn} 
\begin{equation}
\label{tsoft} 
\tilde{Y}_n (x) = \mathrm{P} \exp\Bigl[ig\int^{\infty}_x ds n\cdot A_s (sn)\Bigr]_, 
\end{equation}  
where `P' denotes the path ordering.

After redefining the collinear fields, the soft interaction is decoupled from the collinear interaction \cite{Bauer:2001yt}, 
and the factorized jet cross section in Eq.~(\ref{facjet})
is obtained. The integrated jet function $\mathcal{J}_{n,\Theta} (\mu)$ is defined  with the jet algorithm as
\begin{equation} \label{inunin}
\mathcal{J}_{n,\Theta}  (\mu)= \int dp^2 J_{n,\Theta} (p^2,\mu),
\end{equation}
where $p^2$ is the invariant mass squared of the collinear jet. The unintegrated jet function $J_{n,\Theta}$ is defined as
\begin{equation}
\sum_{X_n} \langle 0| \chi_n^{\alpha} |X_n\rangle \Theta_J \langle X_n | \overline{\chi}_n^{\beta}|0\rangle =
\int \frac{d^4 p_{X_n}}{(2\pi)^3} \frac{\FMslash{n}}{2} \overline{n} \cdot p_{X_n} J_{n,\Theta}  (p_{X_n}^2,\mu) \delta^{\alpha\beta},
\end{equation}
where $\Theta$ specifies the jet algorithm. 
The soft function $\mathcal{S}_{\Theta}$  with the jet algorithm is given by
 \begin{equation}
\mathcal{S}_{\Theta} = \sum_{X_s} \frac{1}{N_c} \mathrm{Tr} \langle 0| \tilde{Y}_{\overline{n}}^{\dagger}
\tilde{Y}_n |X_s\rangle \Theta_{\mathrm{soft}} \langle X_s |\tilde{Y}_n^{\dagger} \tilde{Y}_{\overline{n}} |0\rangle.
\end{equation}
The jet function $\mathcal{J}_{n,\Theta}$ and the soft function $\mathcal{S}_{\Theta}$ are normalized to 1 at tree level, and we compute
them at next-to-leading order with various values of  $\alpha$ in the generalized $k_T$ algorithm.

\section{Jet algorithms\label{jetalgo}}

We now consider the constraints imposed on the phase space by the jet algorithm in Eq.~(\ref{genkt}). At  next-to-leading order (NLO),
there are at most two particles  in a jet. The dominant contribution for the dijet in $e^+ e^-$ annihilation is the
quark-antiquark pair forming a back-to-back jet and a gluon is emitted. We choose the jet in the lightlike
$n$ direction, or in the $\overline{n}$ direction with $n^2 =0$, $\overline{n}^2 =0$ and
$n\cdot \overline{n} =2$.

\begin{figure}[b] 
\begin{center}
\includegraphics[width=5cm]{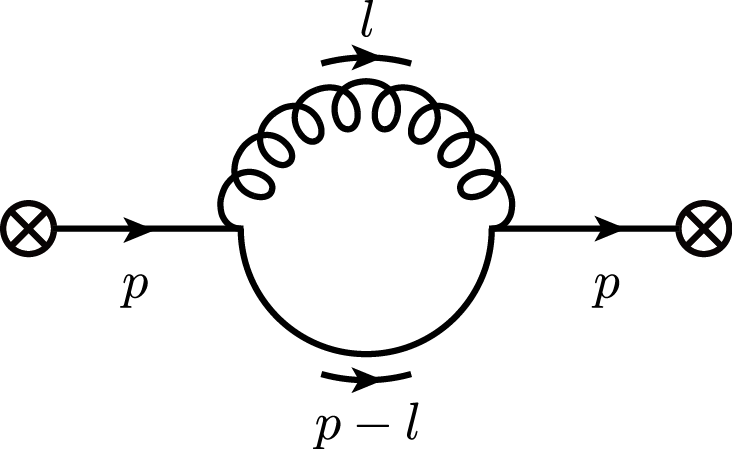}
\end{center}  
\vspace{-0.5cm}
\caption{\baselineskip 3.0ex  Particle configuration and the momentum assignment in constructing the phas space.\label{conf}}
\end{figure}

In applying the jet algorithm, the virtual correction and the real gluon emission are included. These can be
obtained by cutting the diagram in  Fig.~\ref{conf}, which is part of the matrix elements squared for the jet cross
section.  If we cut a single line, it corresponds to the virtual correction. When we cut the loop, there are two particles in 
the final state with momenta $l$ (for a gluon) and $p-l$ (for a quark) in the $n$ direction.  The following nontrivial consideration of the phase
space using the jet algorithm applies to this case with a real gluon emission.
 
The total momentum $p^{\mu}$ of the final state scales as 
\begin{equation}
p^{\mu} = (\overline{n}\cdot p, p_{\perp}, n\cdot p) = (p_-,  p_{\perp}, p_+) \sim 
Q (1,\lambda, \lambda^2),
\end{equation}
where $\lambda$ is the small parameter in SCET.  We choose the $n$ direction as the direction of $p$ such
that $\mathbf{p}_{\perp}=0$, and $p_- =Q$. The collinear gluon momentum $l^{\mu}$
scales as
\begin{equation}
l^{\mu} = (l_- ,  l_{\perp},  l_+) \sim Q (1,\lambda, \lambda^2).
\end{equation}
With this power counting, the collinear momenta of the quark and the gluon can be written as
\begin{equation}
p_q^{\mu} = (Q-l_-, - l_{\perp}, p^2/Q -l_+), \ p_g^{\mu} = (l_-,  l_{\perp}, l_+),
\end{equation}
with their energies
\begin{equation}
E_q = \frac{1}{2} (Q-l_- +p^2/Q -l_+), \ E_g = \frac{1}{2} (l_- + l_+). 
\end{equation}
And the invariant-mass squared $p^2$ is given by
\begin{equation}
p^2 = (p_q + p_g)^2 = \frac{Ql_+}{1-l_-/Q}.
\end{equation}

The phase space constraint from Eq.~(\ref{genkt})  for the jet function changes with the sign of $\alpha$. 
In terms of the gluon momentum $l$ and $Q$, the jet algorithm for $\alpha >0$ is given as
\begin{equation} \label{cocon}
y_{qg} = \left\{
\begin{array}{ll}
\displaystyle \frac{2^{2-\alpha}}{Q^{\alpha}} \frac{l_+ l_-^{\alpha-1}}{(1-l_-/Q)^2} <y_c, &
\displaystyle  0<l_- <\frac{Q}{2}, \\
\displaystyle 2^{2-\alpha} \Bigl(1-\frac{l_-}{Q}\Bigr)^{\alpha -2} \frac{l_+}{l_-} <y_c, & 
\displaystyle\frac{Q}{2} < l_- <Q.
\end{array}
\right.
\end{equation}
The nontrivial jet algorithm appears when the resolution parameter $y_c$ is of order $\lambda^2$ in SCET. Hence we choose $y_c \sim 
\mathcal{O}(\lambda^2)$ from now on. 

\begin{figure}[t] 
\begin{center}
\includegraphics[width=16cm]{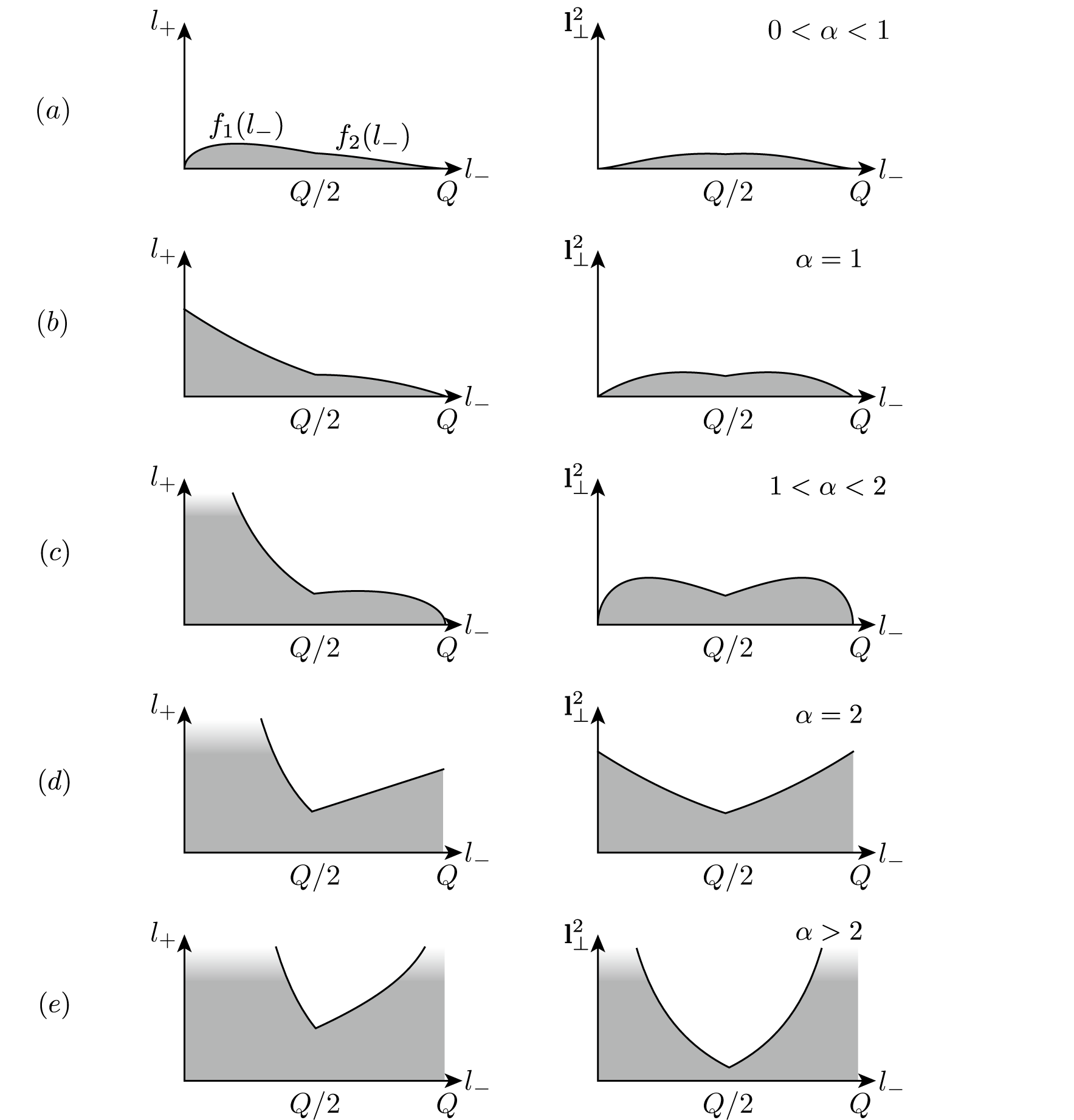}
\end{center}  
\vspace{-0.3cm}
\caption{\baselineskip 3.0ex The  phase space in the general $k_T$ algorithm with $\alpha>0$. The first column shows the 
$(l_-, l_+)$ phase space, while the second column shows the $(l_-, \mathbf{l}_{\perp}^2)$ space.
(a)  $0<\alpha<1$,  (b) $\alpha=1$,  (c)  $1<\alpha<2$,  (d)  $\alpha =2$,  and (e)  $\alpha>2$. 
The functions $f_1$ and $f_2$ describe the boundaries of the phase space
for $0<l_- <Q/2$ and $Q/2<l_- <Q$ respectively, as defined in Eq.~(\ref{f12}).\label{phplus}}
\end{figure}

By letting $l_- = Qx$, $l_+ = \mu y$, the phase space is constrained by\footnote{\baselineskip 3.0ex  The coefficient of the rescaling 
depends on whether the corresponding momentum reaches a
finite value, say, $Q$, or infinity. For the momentum reaching infinity, the dimensionful coefficient is the renormalization scale $\mu$ which
is to be combined with the integral in applying the  dimensional regularization. See the zero-bin contribution or the soft function below.}
\begin{equation} \label{yqg+}
y_{qg} =\left\{ \begin{array}{ll}
\displaystyle 2^{2-\alpha} \frac{\mu}{Q} \frac{yx^{\alpha-1}}{(1-x)^2} < y_c, & 0<x <\frac{1}{2},\\
\displaystyle 2^{2-\alpha} \frac{\mu}{Q} \frac{y(1-x)^{\alpha-2}}{x} < y_c, & \frac{1}{2} <x<1.
\end{array}
\right.
\end{equation}
Solving for $y$, it becomes
\begin{eqnarray} \label{f12}
&& y< 2^{\alpha-2} y_c \frac{Q}{\mu} x^{1-\alpha} (1-x)^2 = f_1 (x),\ \ (0<x <\frac{1}{2}), \ \nonumber \\
&& y< 2^{\alpha-2} y_c \frac{Q}{\mu} x (1-x)^{2-\alpha} = f_2 (x), \ \ (\frac{1}{2} <x<1), 
\end{eqnarray}
where the boundaries are defined in terms of $f_1 (x)$ and $f_2(x)$ respectively. 

The shape of the  phase space in the $(l_-,l_+)$ space with $\alpha>0$ changes drastically for the following five cases: 
(i) $0<\alpha <1$, (ii) $\alpha=1$, (iii)
$1<\alpha<2$,  (iv) $\alpha=2$, (v)  $\alpha>2$. The corresponding phase spaces are shown in Fig.~\ref{phplus}. The figures in the first 
column are the phase spaces in the $(l_-, l_+)$ space, and those in the second column are those in 
the $(l_-, \mathbf{l}_{\perp}^2)$ space. The structure of the
divergence can be inferred from the phase space in the $(l_-, l_+)$ space. But when it becomes ambiguous, the phase space in the 
$(l_-, \mathbf{l}_{\perp}^2)$ space plays a complementary role in deducing the origin of the divergence \cite{Hornig:2009kv,Hornig:2009vb}.
This point will be discussed in detail in Section~\ref{psdiv} after the explicit result is presented.

For $0<\alpha <1$, $l_+$ approaches
zero at the endpoints $l_- =0$ and $Q$.  At $\alpha=1$, $l_+$ becomes a nonzero constant at $l_-=0$, while it
approaches zero at $l_- =Q$. For $1<\alpha <2$, $l_+$ diverges as $l_-$ approaches zero, while it becomes
zero at $l_-=Q$. For $\alpha=2$, $l_+$ diverges at $l_-=0$, while remains finite at $l_-=Q$.  For $\alpha >2$, 
$l_+$ diverges at both endpoints $l_-=0$ and $Q$.  The different behavior of the phase space near the endpoints gives a clue to whether the     
naive contribution to the jet function contains UV or IR divergence. 

In the first column of Fig.~\ref{phplus}, Fig.~\ref{phplus}
(a) and (b) corresponding to $\alpha \le 1$, $l_-$ and $l_+$ never reach infinity, and we expect that the naive collinear contribution has 
IR divergence only. In Fig.~\ref{phplus} (c), (d) and (e) corresponding to $\alpha>1$, $l_+$ can reach infinity and we can expect that 
there may be UV divergence as well as IR divergence when the integration over $l_-$ and $l_+$ is performed. 
However, this is not exactly true. 

In most cases the UV and IR divergences can be expected by looking at the $(l_-, l_+)$ phase space alone. 
But there are some occasions where it is ambiguous. Then we turn to the $(l_-, \mathbf{l}_{\perp}^2)$ phase space. For example, if we consider
the final form of the integral in computing the collinear part with respect to $l_-$ and $\mathbf{l}_{\perp}^2$, 
the second figure in Fig.~\ref{phplus} (c) shows that 
$\mathbf{l}_{\perp}^2$ never reaches infinity, hence there should be only IR divergence for $1<\alpha<2$.  
For $\alpha>2$, both figures in Fig.~\ref{phplus} (e)
indicate that there are both IR and UV divergences. In the second figure in Fig.~\ref{phplus} (d), we expect that there is only IR divergence, 
but the collinear and the soft functions are not well-defined for $\alpha=2$, and it needs more explanation, as we will see below.
 
\begin{figure}[b] 
\begin{center}
\includegraphics[width=11cm]{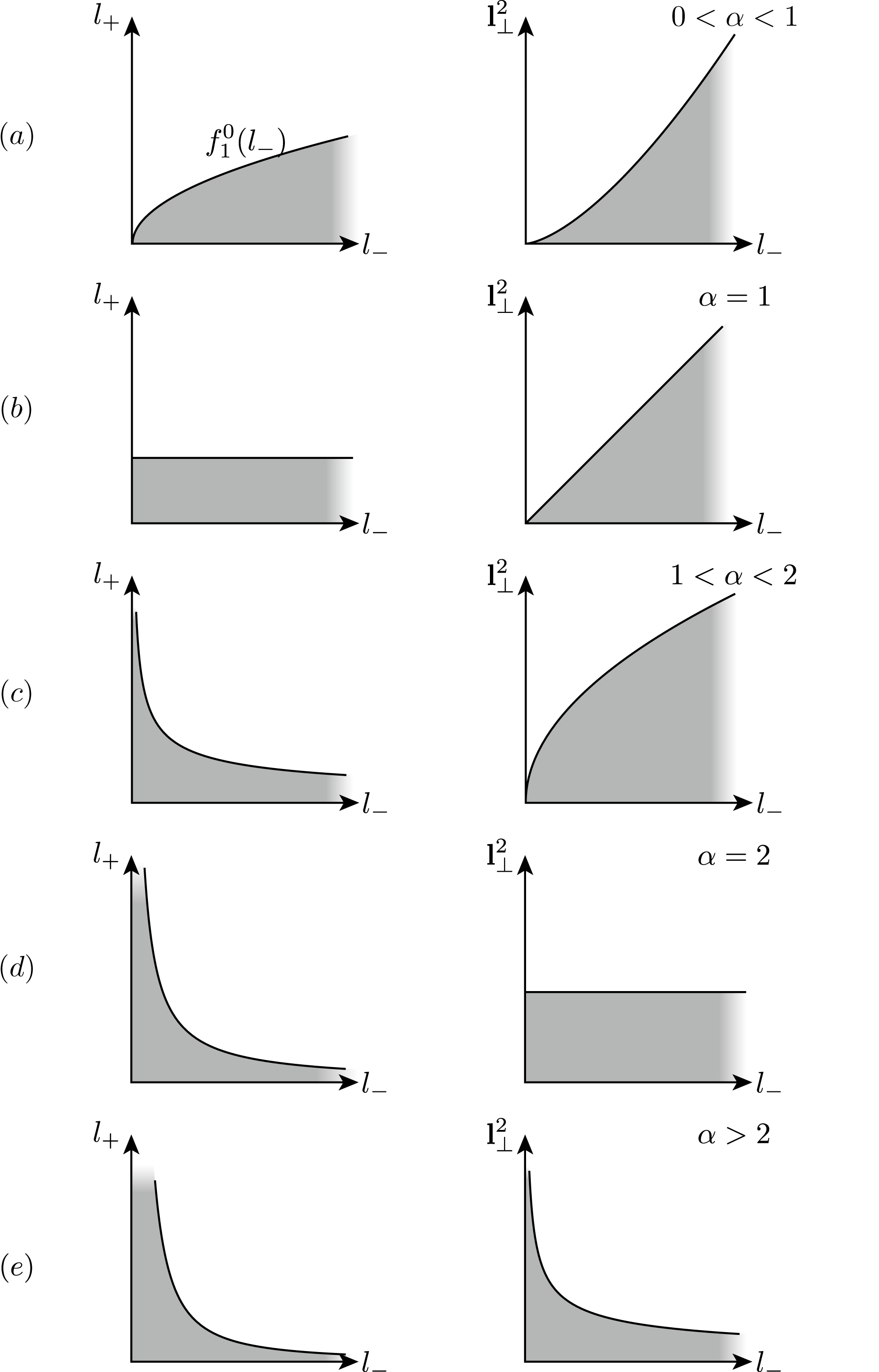}
\end{center}  
\vspace{-0.3cm}
\caption{\baselineskip 3.0ex 
The  phase space for the zero-bin contributions with $\alpha>0$ in the same format as Fig.~\ref{phplus}. 
(a)  $0<\alpha<1$,  (b) $\alpha=1$,  (c)  $1<\alpha<2$,  (d)  $\alpha =2$,  and (e)  $\alpha>2$.
 \label{zeroplus}}
\end{figure}

In order to avoid double counting, we subtract the soft limit from the naive collinear contribution, which is referred to as the zero-bin
subtraction \cite{Manohar:2006nz}.
For the zero-bin contribution, we take the soft limit of the gluon $(l_-, l_{\perp}, l_+) \sim Q(\lambda^2, \lambda^2, \lambda^2)$ and rewrite
the first equation in Eq.~(\ref{f12}). Then the phase space for the zero-bin contribution is given as
\begin{equation}
y_{qg}^0 = 2^{2-\alpha} \frac{l_+ l_-^{\alpha-1}}{Q^{\alpha}}<y_c,
\end{equation}
which can be written, by putting $l_- = \mu x$, $l_+ =\mu y$, as
\begin{equation} \label{f10}
y < 2^{\alpha -2} y_c \Bigl(\frac{Q}{\mu}\Bigr)^{\alpha} x^{1-\alpha}=f_1^0 (x).
\end{equation}
The phase space for the zero-bin contribution corresponding to each case in Fig.~\ref{phplus} is shown in 
Fig.~\ref{zeroplus}. As can be seen clearly in Fig.~\ref{zeroplus}, the zero-bin contribution contains both UV and IR divergences. However, 
the IR divergence is cancelled in the sum of the naive collinear contribution and the zero-bin contribution for $\alpha<2$, 
while the cancellation is incomplete and contains IR divergence for $\alpha>2$.

\begin{figure}[t] 
\begin{center}
\includegraphics[width=11cm]{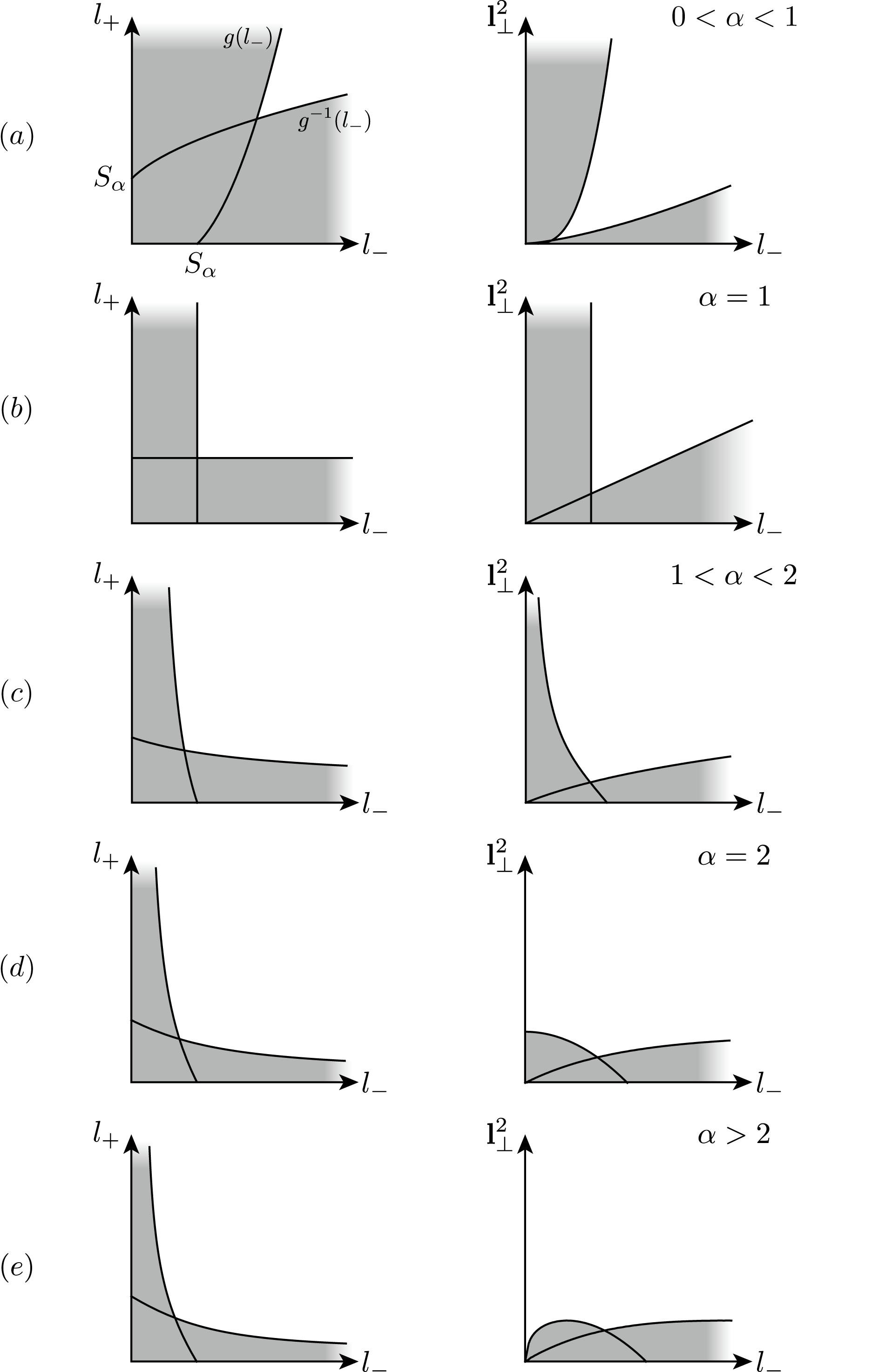}
\end{center}  
\vspace{-0.3cm}
\caption{\baselineskip 3.0ex The  phase space for the soft function with $\alpha>0$ corresponding to 
Fig.~\ref{phplus}.  The jet veto is not 
necessary for $\alpha>0$, and $S_{\alpha} = 2^{(1-\alpha)/\alpha} a^{1/\alpha}$,
(a)  $0<\alpha<1$,  (b) $\alpha=1$,  (c)  $1<\alpha<2$,  (d)  $\alpha =2$,  and (e)  $\alpha>2$.
\label{softplus}}
\end{figure}

For the soft function, $y_{ij}$ satisfies the following relations with the appropriate power counting.
\begin{equation} \label{softij}
y_{qg}^s = \left\{ \begin{array}{ll}
\displaystyle 2^{2-\alpha} \frac{l_+ (l_+ + l_-)^{\alpha-1}}{Q^{\alpha}}< y_c, & n \ \mathrm{jet}, \\
\displaystyle 2^{2-\alpha} \frac{l_- (l_+ + l_-)^{\alpha-1}}{Q^{\alpha}}< y_c, & \overline{n} \ \mathrm{jet}.
\end{array}
\right.
\end{equation}
By replacing $l_- = \mu x$, $l_+ = \mu y$, the boundary from the $\overline{n}$ jet is given by
\begin{equation} \label{softb}
y=g(x) = -x+ \Bigl( \frac{a}{x}\Bigr)^{\frac{1}{\alpha-1}},
\end{equation}
where $a$ is a dimensionless parameter given as
\begin{equation}
a= 2^{\alpha -2}\Bigl( \frac{Q}{\mu}\Bigr)^{\alpha} y_c.
\end{equation}
The boundary from the $n$ jet is given by $y=g^{-1} (x)$.  The phase space for the soft function is shown in 
Fig.~\ref{softplus}.  As can be seen in Fig.~\ref{softplus}, since the region near the origin is totally covered, no jet veto is 
necessary for $\alpha >0$. However,  the jet veto is needed for $\alpha <0$, as in the Sterman-Weinberg jet algorithm \cite{Chay:2015ila}. 

For $\alpha <0$,  the minimum in Eq.~(\ref{genkt}) is switched compared to the case $\alpha>0$. It means 
that the boundary functions are also reversed. That is, the constraint of the phase space for the jet function is 
given by
\begin{eqnarray} \label{f21}
&& y< 2^{\alpha-2} y_c \frac{Q}{\mu} x (1-x)^{2-\alpha} = f_2 (x), \ ( 0<x <\frac{1}{2}), \ \nonumber \\
&&y< 2^{\alpha-2} y_c \frac{Q}{\mu} x^{1-\alpha} (1-x)^2 = f_1 (x), \ (\frac{1}{2} <x<1).
\end{eqnarray}
The phase space for the zero-bin contribution is given by
\begin{equation}
y < 2^{\alpha -2} y_c x,
\end{equation}
and that for the soft function is given by
\begin{eqnarray}
&& y< 2^{\alpha-2} y_c x, \ n\ \mathrm{jet}, \nonumber \\
&& y> \frac{1}{2^{\alpha-2} y_c} x , \ \overline{n} \ \mathrm{jet}, \nonumber \\
&& x+y < 2 \beta \frac{Q}{\mu}, \ \mathrm{jet \ veto}.
 \end{eqnarray}
In contrast to the case with $\alpha>0$, the jet veto is needed here since the phase space near the origin is not completely covered by the 
jet algorithm alone.  And we take $\beta \sim \mathcal{O} (\lambda^2)$ for definite power counting. 

\begin{figure}[t] 
\begin{center}
\includegraphics[width=16cm]{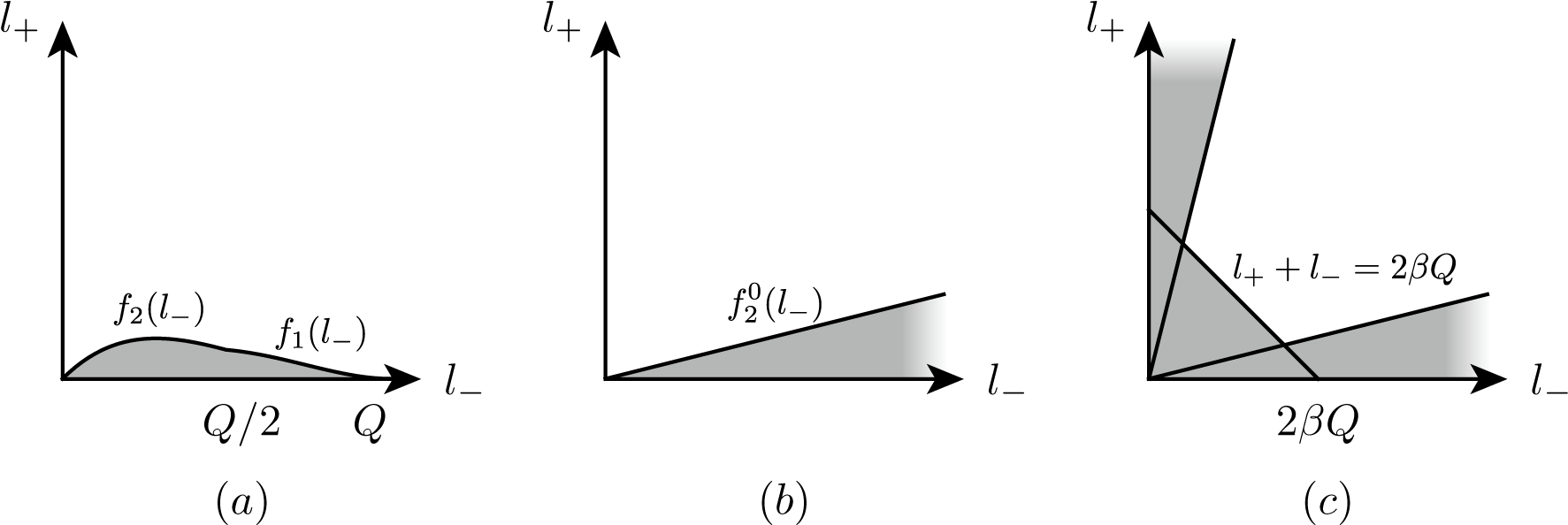}
\end{center}  
\vspace{-0.3cm}
\caption{\baselineskip 3.0ex The  phase space in general $k_T$ algorithm with $\alpha<0$. (a) naive 
collinear contribution  (b) the zero-bin contribution (c)  the soft function.\label{cominus}}
\end{figure}

The phase space for the naive collinear part, the zero-bin contribution, and the soft function for $\alpha<0$ is shown in 
Fig.~\ref{cominus}. The shapes of the phase spaces are similar to the Sterman-Weinberg algorithm or the cone-type algorithm \cite{Chay:2015ila}.
And we expect that the jet and the soft functions are IR finite for all negative values of $\alpha$.

 \section{Generalized $k_T$ jet algorithm with $\alpha >0$\label{posalp}}

\subsection{Jet function}

\begin{figure}[b] 
\begin{center}
\includegraphics[width=16cm]{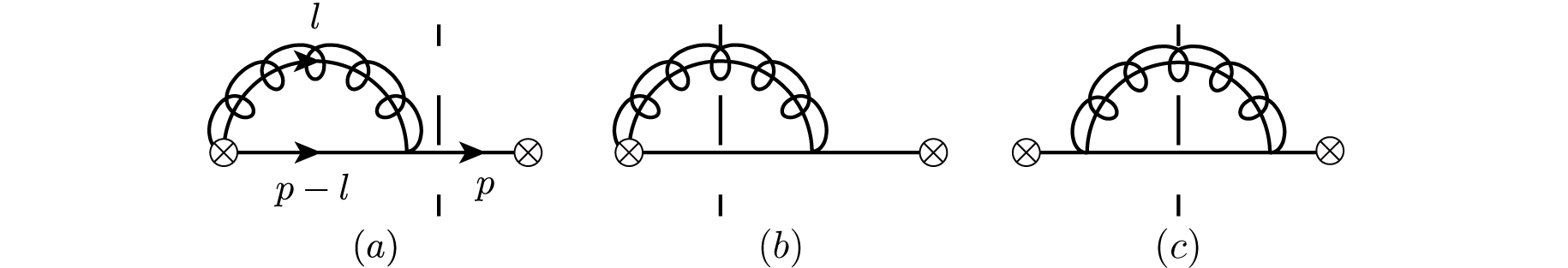}
\end{center}  
\vspace{-0.3cm}
\caption{\baselineskip 3.0ex Feynman diagrams for the jet  function at one loop  
(a) virtual correction (b) real gluon emission from the Wilson line (c) real gluon emission.\label{intjet}}
\end{figure}

The Feynman diagrams for the jet function in the $n$ direction, are shown in Fig.~\ref{intjet}. The dashed line 
is the cut, therefore Fig.~\ref{intjet} (a) is the virtual correction, while Fig.~\ref{intjet} (b) and (c) 
are the real contributions. The mirror images of (a) and (b)  are  omitted in Fig.~\ref{intjet}, but they are 
included in the computation. Each diagram is accompanied by the corresponding zero-bin contribution, which
should be subtracted to obtain the collinear jet function. 

The dimensional regularization states that 
\begin{equation}
\mu^{\eps}\int_0^{\infty} dl\, l^{-1-\eps} = \frac{1}{\euv} -\frac{1}{\eir},
\end{equation}
where $l$ is a momentum variable. The integral vanishes identically if we do not distinguish the 
UV and IR poles since the integral is a scaleless integral. However, we distinguish the UV and IR poles since we are interested in identifying the
sources of the divergence.
 
The virtual correction in Fig.~\ref{intjet} (a) is independent of the jet algorithms and
the net collinear contribution $M_a$ is given by the naive collinear contribution $\tilde{M}_a$,  subtracted by 
the zero-bin contribution $M_a^0$:
\begin{eqnarray}
\tilde{M}_a &=& \frac{\alpha_s C_F}{2\pi} \Bigl( \frac{1}{\euv} -\frac{1}{\eir}\Bigr) \Bigl( \frac{1}{\eir} +1 
+\ln \frac{\mu}{Q}\Bigr), 
\nonumber \\
M_a^0 &=& -\frac{\alpha_s C_F}{2\pi} \Bigl( \frac{1}{\euv} -\frac{1}{\eir}\Bigr)^2,  \nonumber \\
M_a &=&  \tilde{M}_a - M_a^0 =  \frac{\alpha_s C_F}{2\pi} 
\Bigl( \frac{1}{\euv} -\frac{1}{\eir}\Bigr) \Bigl( \frac{1}{\euv} +1 +\ln \frac{\mu}{Q}\Bigr).
\end{eqnarray}

\subsubsection{The case $0<\alpha < 1$}

The naive collinear contribution for $0<\alpha<1$ in Fig.~\ref{intjet} (b) is given as
\begin{eqnarray} \label{naimb}
\tilde{M}_b &=&  \frac{\alpha_s C_F}{2\pi} \frac{e^{\gamma_{\mathrm{E}}\eps}}{\Gamma (1-\eps)}  \Bigl(\frac{\mu}{Q}\Bigr)^{\eps}
\Bigl[\int_0^{1/2} dx x^{-1-\eps} (1-x)  \int_0^{f_1 (x)} dy y^{-1-\eps} \nonumber \\
&&
+ \int_{1/2}^1 dx x^{-1-\eps} (1-x) 
\int_0^{f_2 (x)} dy y^{-1-\eps} \Bigr]\nonumber \\
&=& \frac{\alpha_s C_F}{2\pi}  \Bigl[ \frac{1}{\alpha-2}\Bigl( -\frac{1}{\eir^2} -\frac{1}{\eir} \ln \frac{\mu^2}{2^{\alpha-2} y_c Q^2 } 
- \frac{1}{2}\ln^2 \frac{\mu^2}{2^{\alpha-2} y_c Q^2} +\frac{\pi^2}{12}\Bigr) \nonumber \\
&& +\frac{1}{\eir}  +  \ln \frac{\mu^2}{2^{\alpha-2} y_c Q^2 } + (4-\alpha) \Bigl(1-\frac{\pi^2}{12}\Bigr) -\alpha \ln 2 \Bigr].
\end{eqnarray}
This explicit calculation supports the expectation from  the phase space in Fig.~\ref{phplus} (a) that the poles are of the IR origin. 
This can be also seen by the fact  that the integral of $\tilde{M}_b$  in Eq.~(\ref{naimb}) exists only when $\eps<0$. 

The corresponding zero-bin contribution is given by
\begin{equation}
M_b^0 = \frac{\alpha_s C_F}{2\pi} \frac{e^{\gamma_{\mathrm{E}}\eps}}{\Gamma (1-\eps)} \int_0^{\infty} 
dx x^{-1-\eps} \int_0^{f_1^0 (x)} dy 
y^{-1-\eps},
\end{equation}
where $f_1^0 (x)$ is the zero-bin limit of $f_1 (x)$ and is given in Eq.~(\ref{f10}).
From Fig.~\ref{zeroplus}, since $l_-$, $l_+$ can both approach zero and infinity, the zero-bin contribution contains both UV and IR divergences. 
In order to separate the UV and IR contributions,  the integral is decomposed to be written as
\begin{eqnarray}
M_b^0 &=& \frac{\alpha_s C_F}{2\pi} \frac{e^{\gamma_{\mathrm{E}}\eps}}{\Gamma (1-\eps)}  \Bigl[ 
\int_0^{\eta} dx x^{-1-\eps} \int_0^{f_1^0 (x)} dy y^{-1-\eps} \nonumber \\
&&
+ \int_{\eta}^{\infty} dx x^{-1-\eps} \Bigl( \int_0^{\infty} dy y^{-1-\eps} -\int_{f_1^0 (x)}^{\infty} dy 
y^{-1-\eps}\Bigr) \Bigr] \nonumber \\
&=& \frac{\alpha_s C_F}{2\pi} \Bigl(\frac{1}{\euv} -\frac{1}{\eir}\Bigr) \Bigl[ \frac{1}{\alpha-2} \Bigl( 
\frac{1}{\euv} +\frac{1}{\eir} + \ln \frac{\mu^2}{2^{\alpha-2} y_c Q^2} \Bigr)  +\frac{1}{\euv}  + \ln \frac{\mu}{Q} \Bigr],
\end{eqnarray}
where $\eta$ is a small positive number to separate the IR and UV regions. The final result is independent of $\eta$.
This decomposition looks complicated, but it is arranged in such a way that the pole from the integral 
including the infinity (zero) is of the UV (IR) origin with $\eps>0$ ($\eps<0$) \cite{Chay:2015ila}.

And the net collinear contribution from Fig.~\ref{intjet} (b) is given by
\begin{eqnarray}
M_b &=& \tilde{M_b} -M_b^0   \\
&=& \frac{\alpha_s C_F}{2\pi} \Bigl[ \frac{1}{\alpha-2} \Bigl( -\frac{1}{\euv^2}
 -\frac{1}{\euv}  \ln \frac{\mu^2}{2^{\alpha-2} y_c Q^2} -\frac{1}{2}  \ln^2 \frac{\mu^2}{2^{\alpha-2} y_c Q^2}
+\frac{\pi^2}{12}\Bigl) \nonumber \\
& &+   \frac{1}{\eir} +\ln \frac{\mu^2}{2^{\alpha-2}y_c Q^2} -\Bigl(\frac{1}{\euv} -\frac{1}{\eir}\Bigr) \Bigl( \frac{1}{\euv} +\ln \frac{\mu}{Q}
\Bigr)  +(4-\alpha) \Bigl( 1-\frac{\pi^2}{12}\Bigr) -\alpha \ln 2\Bigr]. \nonumber 
\end{eqnarray}

The naive collinear contribution from Fig.~\ref{intjet} (c) is given by
\begin{eqnarray} \label{mcal+}
\tilde{M_c} &=&  \frac{\alpha_s C_F}{2\pi} \frac{(1-\eps) e^{\gamma_{\mathrm{E}}\eps}}{\Gamma (1-\eps)}  
\Bigl( \frac{\mu}{Q}\Bigr)^{\eps}
\Bigl[ \int_0^{1/2} dx x^{1-\eps} \int_0^{f_1 (x)} dy y^{-1-\eps} + \int_{1/2}^1 dx x^{1-\eps} \int_0^{f_2 (x)} 
dy y^{-1-\eps} \Bigr] \nonumber \\
&=& \frac{\alpha_s C_F}{2\pi} \Bigl( -\frac{1}{2\eir} -\frac{1}{2} \ln \frac{\mu^2}{2^{\alpha-2} y_c Q^2} +\frac{\alpha-3}{2}  
+\frac{\alpha}{2}\ln 2  \Bigr),
\end{eqnarray}
while the zero-bin contribution is subleading and suppressed.  
Adding all the terms with the wavefunction renormalization and the residue at one loop
\begin{equation}
Z_{\xi}^{(1)} +R_{\xi}^{(1)} = -\frac{\alpha_s C_F}{2\pi} \Bigl( \frac{1}{2\euv} -\frac{1}{2\eir}\Bigr),
\end{equation}
the collinear contribution for $0<\alpha < 1$ at order $\alpha_s$ is IR finite and is given as
\begin{eqnarray} \label{mb1}
M_{\mathrm{coll}}^{0<\alpha<1} &=& 2 (M_a + M_b) + \tilde{M}_c +Z_{\xi}^{(1)} +R_{\xi}^{(1)}  \nonumber \\
&=&\frac{\alpha_s C_F}{2\pi} \Bigl[ \frac{1}{\alpha-2} \Bigl( -\frac{2}{\euv^2} -\frac{2}{\euv}  \ln 
\frac{\mu^2}{2^{\alpha-2} y_c Q^2} - \ln^2 \frac{\mu^2}{2^{\alpha-2} y_c Q^2} +\frac{\pi^2}{6}\Bigr) \nonumber \\
&& +\frac{3}{2} \Bigl(\frac{1}{\euv} +\ln \frac{\mu^2}{2^{\alpha-2} y_c Q^2} \Bigr)   +\frac{13-3\alpha}{2}  
-\frac{3\alpha}{2}\ln 2  +(\alpha-4)\frac{\pi^2}{6}  \Bigr]. 
\end{eqnarray}
By removing the UV poles, the jet function for $0<\alpha<1$ at order $\alpha_s$ is given by
\begin{eqnarray}
\mathcal{J}_n^{(1)} (Q, y_c,\mu) &=& \frac{\alpha_s C_F}{2\pi}  \Bigl[\frac{1}{\alpha-2} \Bigl( - \ln^2 \frac{\mu^2}{2^{\alpha-2} y_c Q^2}
 +\frac{\pi^2}{6}\Bigr) +  \frac{3}{2}  \ln \frac{\mu^2}{2^{\alpha-2} y_c Q^2} \nonumber \\
&&+\frac{13-3\alpha}{2}  -\frac{3\alpha}{2}\ln 2  +(\alpha-4)\frac{\pi^2}{6}  \Bigr].
\end{eqnarray}

\subsubsection{The case $\alpha=1$}

Note that the phase space for $\alpha=1$ (naive and zero-bin) allows the same method of  integration with the same 
specification of the poles as in the case $0<\alpha <1$. The only difference is that $l_+$ approaches a nonzero, finite value as $l_-$ 
approaches zero in the phase space for the naive collinear contribution, hence the case 
$\alpha =1$ can be included in Eq.~(\ref{mb1}). The case $\alpha=1$ is very similar to the JADE jet algorithm as far as the phase space is 
concerned. 

In the JADE algorithm, the invariant mass $M_{ik}^2$ of every pair of the final-state partons $i$ and $k$ should be less than $jQ^2$, where $j$
is some fraction (of order $\lambda^2$ in the context of SCET) to form a jet. At next-to-leading order,  the 
JADE algorithm is written as
\begin{eqnarray}
l_+ < j (Q-l_- ), && \ \ \text{collinear}, \nonumber \\
 l_+ < jQ, && \ \ \text{zero-bin}, \nonumber \\
 l_- <jQ, \ l_+ <jQ,  && \ \ \text{soft}.
\end{eqnarray}

If we compare this phase space with that of Fig.~\ref{phplus} (b), \ref{zeroplus} (b) and \ref{softplus} (b) for $\alpha=1$, the basic structure 
near the endpoints of the phase space is the same, and the only difference lies in the middle region of Fig.~\ref{phplus} (b). Therefore we expect
that the divergence structure for the case $\alpha=1$ and the JADE algorithm is the same except the finite part in the dijet cross section. 
In fact, performing the calculation using the JADE algorithm, the collinear part is given as
\begin{equation} \label{jadejet}
M_{\mathrm{coll}}^{\mathrm{JADE}} 
 = \frac{\alpha_s C_F}{2\pi}\Bigl[\frac{2}{\euv^2} + \frac{1}{\euv} \Bigl( \frac{3}{2} +2 \ln \frac{\mu^2}{jQ^2} \Bigr) 
+\frac{3}{2}\ln \frac{\mu^2}{jQ^2} + \ln^2 \frac{\mu^2}{jQ^2} +\frac{7}{2} -\frac{\pi^2}{2}\Bigr],
\end{equation}
which is consistent with the result in Ref.~\cite{Cheung:2009sg}. And the corresponding collinear part for $\alpha =1$ 
from Eq.~(\ref{mb1}) is given by
\begin{equation} \label{jetalpha1}
M_{\mathrm{coll}}^{\alpha=1}  
=\frac{\alpha_s C_F}{2\pi}  \Bigl[ \frac{2}{\euv^2} +\frac{1}{\euv}\Bigl( \frac{3}{2} +2\ln \frac{2\mu^2}{y_c Q^2}\Bigr)
+\frac{3}{2}  \ln \frac{2\mu^2}{y_c Q^2} + \ln^2 \frac{2\mu^2}{y_c Q^2} +5- \frac{3}{2}\ln 2 -\frac{2\pi^2}{3}\Bigr].
\end{equation}
If we put $j=y_c/2$, all the UV divergence and the logarithmic terms are reproduced except the finite terms, which is why we say that the 
generalized jet algorithm with $\alpha=1$ is basically identical to the JADE algorithm. 

From Eqs.~(\ref{jadejet}) and (\ref{jetalpha1}), the jet functions in the JADE algorithm and in the case with $\alpha=1$ are given as
\begin{eqnarray}
\mathcal{J}^{(1)}_{n,\mathrm{JADE}} (Q,j,\mu) &=& \frac{\alpha_s C_F}{2\pi}\Bigl(
\frac{3}{2}\ln \frac{\mu^2}{jQ^2} + \ln^2 \frac{\mu^2}{jQ^2} +\frac{7}{2} -\frac{\pi^2}{2}\Bigr), \nonumber \\
\mathcal{J}^{(1)}_{n,\alpha=1} (Q,y_c,\mu) &=& \frac{\alpha_s C_F}{2\pi}  \Bigl(
\frac{3}{2}  \ln \frac{2\mu^2}{y_c Q^2} + \ln^2 \frac{2\mu^2}{y_c Q^2} +5- \frac{3}{2}\ln 2 -\frac{2\pi^2}{3}\Bigr).
\end{eqnarray}

 \subsubsection{The case $1< \alpha< 2$}
 By looking at the phase space for the naive collinear contribution in Fig.~\ref{phplus} (c) in the $(l_-, l_+)$ space, one might naively expect
that there could be a 
mixed UV and IR divergence since $l_+$ reaches infinity as $l_-$ approaches zero. However it becomes clear when we look at the phase space in the 
$(l_-, \mathbf{l}_{\perp}^2)$ space. It means that, when we perform the integral over $l$, the $l_+$ is first performed and the remaining 
integral is to be performed with respect to $l_-$ and $\mathbf{l}_{\perp}^2$ \cite{Hornig:2009kv,Hornig:2009vb}.
Then the divergent behavior is determined by $\mathbf{l}_{\perp}^2 = l_+l_-$. 
If the divergence occurs when $\mathbf{l}_{\perp}^2\rightarrow \infty$, it is the UV divergence. On the other 
hand, if the divergence occurs  when $\mathbf{l}_{\perp}^2\rightarrow 0$, it is the IR divergence. 
The constraint for $\mathbf{l}_{\perp}^2$ from the jet algorithm is given by
\begin{equation}
\mathbf{l}_{\perp}^2 = l_+ l_- < \left\{ \begin{array}{ll}
2^{\alpha -2} Q\mu y_c x^{2-\alpha} (1-x)^2, & 0<x<\frac{1}{2}, \\
2^{\alpha-2} Q\mu y_c x^2 (1-x)^{2-\alpha}, & \frac{1}{2} <x<1.
\end{array}
\right.
\end{equation}
which approaches zero ($1<\alpha<2$) as $x\rightarrow 0$. It can be also seen 
in Fig.~\ref{phplus} (c) with the plot in the  $(l_-,\mathbf{l}_{\perp}^2)$ space,  $\mathbf{l}_{\perp}^2$ never approaches infinity. 
Therefore the poles should be of the IR origin.

The collinear contributions computed for the case $0<\alpha \le 1$ can be exactly performed with 
the same prescription on the signs of $\eps$, hence giving the same result for $1<\alpha<2$.  It also turns out that 
\begin{equation}
\lim_{\alpha \rightarrow 1_-} M_{\mathrm{coll}}^{0<\alpha<1} =\lim_{\alpha \rightarrow 1_+} M_{\mathrm{coll}}^{1<\alpha<2}.
\end{equation}
Therefore Eq.~(\ref{mb1}) can be extended to the case $1< \alpha< 2$, and it is continuous at $\alpha=1$. 
 
\subsubsection{The case $\alpha>2$}
The situation is completely different for $\alpha>2$. In the naive collinear and the zero-bin contributions, the 
phase space shows that  $\mathbf{l}_{\perp}^2$ can reach zero or infinity. Therefore we expect that the 
collinear contributions contain both UV and IR poles, which makes the factorization obsolete because the jet 
function and the soft function are not IR finite. However, as will be seen later, the sum
of  the jet functions and the soft function is IR finite.

In the real gluon emission, the expression for $\tilde{M}_b$ is given by
\begin{eqnarray}  \label{nmb2}
\tilde{M}_b &=&  \frac{\alpha_s C_F}{2\pi} \frac{e^{\gamma_{\mathrm{E}}\eps}}{\Gamma (1-\eps)} \Bigl(\frac{\mu}{Q}\Bigr)^{\eps}
\Bigl[\int_0^{1/2} dx x^{-1-\eps} (1-x) \int_0^{f_1 (x)} dy y^{-1-\eps} \nonumber \\
&&+ \int_{1/2}^1 dx x^{-1-\eps} (1-x) 
\int_0^{f_2 (x)} dy y^{-1-\eps} \Bigr]. 
\end{eqnarray}
But the function $f_1(x)$
describing the boundary of the phase space for $0<l_- <Q/2$ diverges as $l_-$ approaches zero by noting that 
the function $f_1 (x)$ is proportional to $x^{1-\alpha} (1-x)^2$.  It requires the modification of the first integral
in Eq.~(\ref{nmb2}) as
\begin{eqnarray}
&&\int_0^{1/2} dx x^{-1-\eps} (1-x) \int_0^{f_1 (x)} dy y^{-1-\eps}\nonumber \\
&=& \int_0^{1/2} dx x^{-1-\eps} (1-x) \Bigl(\int_0^{\infty} dy y^{-1-\eps}- \int_{f_1 (x)}^{\infty} dy 
y^{-1-\eps} \Bigr) \nonumber \\
&=&  \int_0^{1/2} dx x^{-1-\eps} (1-x) \Bigl(\frac{1}{\euv}-\frac{1}{\eir}- \frac{1}{\euv}
\Bigl(f_1 (x)\Bigr)^{-\eps}\Bigr). 
\end{eqnarray}
This manipulation is necessary since the remaining integral converges for $\eps>0$, hence the integral over 
$y$ should produce $\euv$.  The boundary of the phase space, $f_2 (x)$ for $Q/2 < l_- <Q$ is proportional to 
$x (1-x)^{2-\alpha}$, and it also diverges as $l_-$ approaches $Q$ ($x\rightarrow 1$). However, the integrand is 
proportional to $1-x$, and there is no UV divergence involved. The naive collinear contribution is given as
\begin{eqnarray}
\tilde{M}_b &=& \frac{\alpha_s C_F}{2\pi}\Bigl[ \frac{1}{\alpha-2} \Bigl( -\frac{1}{\euv^2} -\frac{1}{\euv}  
\ln \frac{\mu^2}{2^{\alpha-2} y_c Q^2} 
-\frac{1}{2}  \ln^2 \frac{\mu^2}{2^{\alpha-2} y_c Q^2} +\frac{\pi^2}{12}\Bigr) \nonumber \\
&+& \frac{1}{\eir^2} -\frac{1}{\euv\eir} +\frac{1}{\eir} + \Bigl( \frac{1}{\eir} -\frac{1}{\euv}\Bigr)  \ln \frac{\mu}{Q }   
+  \ln \frac{\mu^2}{2^{\alpha-2} y_c Q^2} \nonumber \\
&+& 2-2\ln 2 -\frac{\pi^2}{6} -(\alpha-2)\Bigl( 1+  \ln 2 -\frac{\pi^2}{12}\Bigr)\Bigr].
\end{eqnarray}

In the zero-bin contribution, a similar modification of the integral is necessary.  It is given as
\begin{eqnarray}
M_b^0 &=&   \frac{\alpha_s C_F}{2\pi} \frac{e^{\gamma_{\mathrm{E}}}}{\Gamma (1-\eps)}  \Bigl( 
\frac{\mu}{Q}\Bigr)^{\eps}
\int_0^{\infty} dx x^{-1-\eps} \int_0^{f_2^0 (x)} dy y^{-1-\eps} \nonumber \\
&=&\frac{\alpha_s C_F}{2\pi} \frac{e^{\gamma_{\mathrm{E}}}}{\Gamma (1-\eps)} \Bigl( 
\frac{\mu}{Q}\Bigr)^{\eps} 
\Bigl( \int_0^{\eta} dx x^{-1-\eps} \int_0^{f_0 (x)} dy y^{-1-\eps} + \int_{\eta}^{\infty} dx x^{-1-\eps} 
\int_0^{f_0 (x)} 
dy y^{-1-\eps} \Bigr) \nonumber \\
&=& \frac{\alpha_s C_F}{2\pi} \frac{e^{\gamma_{\mathrm{E}}}}{\Gamma (1-\eps)}  \Bigl( 
\frac{\mu}{Q}\Bigr)^{\eps}
\Bigl[\Bigl( \frac{1}{\euv} -\frac{1}{\eir} \Bigr) 
\int_0^{\eta}  dx x^{-1-\eps} -\int_0^{\eta} x^{-1-\eps}  \int_{f_0 (x)}^{\infty} dy y^{-1-\eps}  \nonumber \\
&&+ \int_{\eta}^{\infty} dx x^{-1-\eps} \int_0^{f_0 (x)} dy y^{-1-\eps} \Bigr] \nonumber \\
&=& \frac{\alpha_s C_F}{2\pi} \Bigl[ \frac{-1}{\alpha-2} \Bigl( \frac{1}{\euv} -\frac{1}{\eir}\Bigr) 
\Bigl(\frac{1}{\euv} +\frac{1}{\eir} 
+\ln \frac{\mu^2}{2^{\alpha-2} y_c Q^2} \Bigr) +\frac{1}{\eir^2} -\frac{1}{\euv\eir} \nonumber \\
&&  - \Bigl( \frac{1}{\euv} -\frac{1}{\eir}\Bigr) \ln \frac{\mu}{Q}   \Bigr].
\end{eqnarray}
The net collinear contribution is given by
\begin{eqnarray}
M_b &=& \tilde{M}_b -M_b^0 \nonumber  \\
&=& \frac{\alpha_s C_F}{2\pi}  \Bigl[ \frac{1}{\alpha -2} \Bigl(-\frac{1}{\eir^2} -\frac{1}{\eir} \ln 
\frac{\mu^2}{2^{\alpha-2} y_c Q^2} 
-\frac{1}{2} \ln^2 \frac{\mu^2}{2^{\alpha-2} y_c Q^2} +\frac{\pi^2}{12}\Bigl)  \nonumber \\
&& +\frac{1}{\eir}  +  \ln \frac{\mu^2}{2^{\alpha-2} y_c Q^2}
+4-\alpha -\alpha \ln 2 +(\alpha-4)  \frac{\pi^2}{12}  \Bigr]. 
\end{eqnarray}

Fig.~\ref{intjet} (c)   yields the same results as in the case $0<\alpha<2$.  Summing all the contributions, the 
net collinear part is given as
\begin{eqnarray} \label{co2l}
M_{\mathrm{coll}}^{\alpha>2} &=& \frac{\alpha_s C_F}{2\pi}  \Bigl[ \frac{1}{\alpha-2}\Bigl( -\frac{2}{\eir^2} 
-\frac{2}{\eir}  \ln \frac{\mu^2}{2^{\alpha -2} y_c Q^2}
- \ln^2 \frac{\mu^2}{2^{\alpha -2} y_c Q^2} +\frac{\pi^2}{6}\Bigr)\nonumber \\
&& +\frac{2}{\euv^2} -\frac{2}{\euv\eir} +\frac{3}{2\euv}    +\Bigl(\frac{1}{\euv}-\frac{1}{\eir} \Bigr) \ln \frac{\mu^2}{Q^2}
+ \frac{3}{2}  \ln \frac{\mu^2}{2^{\alpha -2} y_c Q^2} \nonumber \\
&&+ \frac{13-3\alpha}{2}-\frac{3\alpha}{2} \ln 2 + (\alpha-4)  \frac{\pi^2}{6}\Bigr].
\end{eqnarray}
In this case, the collinear part is both UV and IR divergent. Therefore it does not have any 
physical meaning for $\alpha> 2$. Compared with the collinear part for $0<\alpha<2$ given in Eq.~(\ref{mb1}) , if we put $\euv=\eir=\eps$,
the result is the same. However, we stress that this is misleading because the collinear part for $0<\alpha <2$ is IR finite and there are only UV poles,
while that for $\alpha>2$ contains both UV and IR divergences. In this case, simply putting $\euv=\eir=\eps$ ruins the physical implication on 
the divergences. But, as we will see later,  the sum of the IR divergent collinear part and the soft part, which is 
also IR divergent, is IR finite. Therefore  the combined collinear and soft part is physically 
well-defined, but the factorization is not realized in a strict sense.

\subsubsection{The case $\alpha=2$}
The jet algorithm with $\alpha=2$ corresponds to the exclusive $k_T$ algorithm. A serious trouble in this case is that the jet function has a pole 
at $\alpha=2$ and is not well-defined. This is to be contrasted with the case $\alpha=1$. For $\alpha=1$, the structure of the phase space 
remains continuous for $0<\alpha<1$ and $1<\alpha <2$, hence the same divergence structure follows. However, for $\alpha=2$, the shape
of the phase space abruptly changes from $0<\alpha<1$ to $\alpha>2$. The structure of the divergence also abruptly changes accordingly.
There is only IR divergence for $0<\alpha<1$, while there are UV, IR divergences for $\alpha>2$. Therefore the jet function for $\alpha=2$ 
cannot be obtained by taking the limiting case as $\alpha \rightarrow 2$ from either side. This is manifested by the existence of the pole in 
$1/(\alpha-2)$. 

However, as we will see later, the dijet cross section in the $k_T$ algorithm is finite, but we emphasize that the factorization
breaks down for $\alpha=2$.  Since the behavior of the jet function, as well as the soft function, is peculiar, it is worth mentioning 
what happens at $\alpha=2$ in detail later.

\begin{figure}[b] 
\begin{center}
\includegraphics[width=16cm]{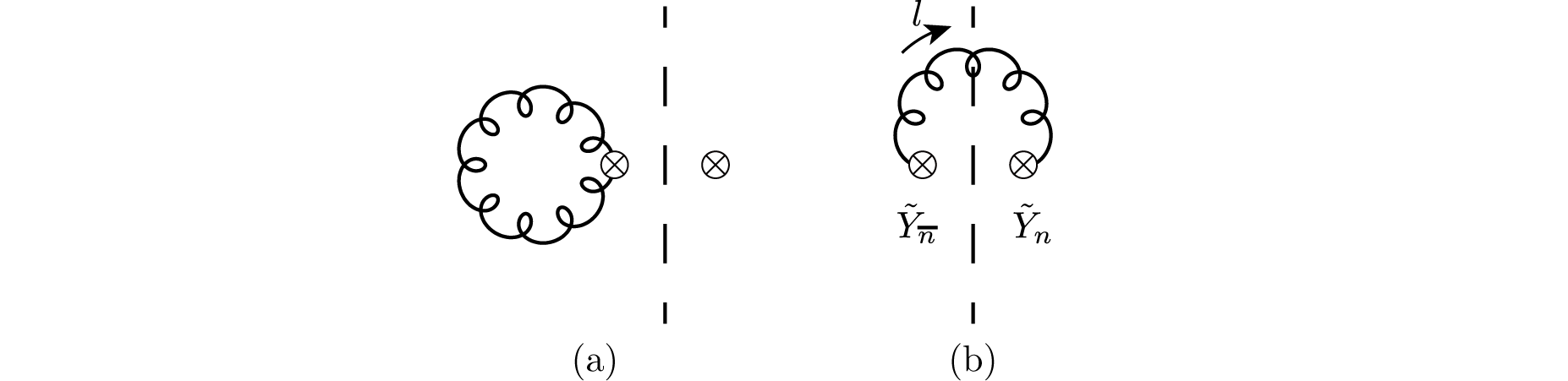}
\end{center}  
\vspace{-0.3cm}
\caption{\baselineskip 3.0ex Feynman diagrams for the soft  function at one loop  
(a) virtual corrections (b) real gluon emission.\label{feynsoft}}
\end{figure}

\subsection{Soft function}
The Feynman diagrams for the soft function at one loop is shown in Fig.~\ref{feynsoft}, in which the hermitian
conjugate is omitted.
The phase space for the soft function is shown in Fig.~\ref{softplus}. The prescription for the UV and IR 
divergences is the same as in the case of the jet function according to the possible values  $\mathbf{l}_{\perp}^2$.

The virtual correction in Fig.~\ref{feynsoft} (a) is given as
\begin{equation}
S_a = -\frac{\alpha_s C_F}{2\pi}\Bigl(\frac{1}{\euv} -\frac{1}{\eir}\Bigr)^2,
\end{equation} 
which is independent of the jet algorithm. We compute the real gluon emission from Fig.~\ref{feynsoft} (b) for various values of $\alpha>0$.

\subsubsection{The case $0<\alpha<1$}

The structure of the phase space for the soft function is shown in Fig.~\ref{softplus} (a), and it is apparent that both the UV and IR divergences
may exist. The contribution of the real gluon emission is given by
\begin{equation}
S_b = 2\frac{\alpha_s C_F}{2\pi}  \frac{e^{\gamma_{\mathrm{E}}\eps}}{\Gamma (1-\eps)}  \Bigl[ \int_0^A dx x^{-1-\eps} \int_x^{\infty}
dy y^{-1-\eps} +\int_A^{\infty} dx x^{-1-\eps} \int_{g(x)}^{\infty} dy y^{-1-\eps}\Bigr],
\end{equation}
where $A= (2^{1-\alpha} a)^{1/\alpha}$ is the point where two boundaries meet, and $g(x)$ describes the boundary for the $\overline{n}$ jet, and
is given in Eq.~(\ref{softb}). We divide the phase space symmetrically with respect to the line $y=x$, and compute only the upper half, 
hence the factor 2 is multiplied. By carefully separating the UV and IR divergences, the amplitude for the soft function, including the
hermitian conjuagates, is given by
\begin{eqnarray}
M_{\mathrm{soft}}^{0<\alpha<1} &=&  2(S_a + S_b)  \\
&=&\frac{\alpha_s C_F}{2\pi}  \Bigl[ \frac{1}{\alpha-2} \Bigl( \frac{4}{\euv^2} -\frac{4}{\euv} \ln a
 + 2\ln^2  a  -\frac{\pi^2}{3}\Bigr) + \frac{2}{\euv^2} -\frac{2}{\alpha} \ln^2 a  
 +\pi^2 \Bigl( \frac{1}{3\alpha} -\frac{1}{2} \Bigr) \Bigr], \nonumber
\end{eqnarray}
where $a= 2^{\alpha-2} y_c (Q/\mu)^{\alpha}$. The soft function at one loop for $0< \alpha <1$ is given as
\begin{equation}
\mathcal{S}^{(1)} (Q, y_c, \mu) = \frac{\alpha_s C_F}{2\pi} \Bigl[\frac{4}{\alpha(\alpha-2)}  \ln^2  2^{\alpha-2} y_c \Bigl(\frac{Q}{\mu}\Bigr)^{\alpha}  -\pi^2 \Bigl( \frac{2}{3\alpha(\alpha-2)} +\frac{1}{2} \Bigr) \Bigr]. 
\end{equation}

The IR divergence from the virtual correction is cancelled by that in the real gluon emission, and there remains only UV divergence. This fact 
can be seen transparently from the argument of the phase space \cite{Hornig:2009kv}. The virtual contribution corresponds to the same integral
as for the real gluon emission, but the integration region covers all the phase space, with a minus sign. Therefore the total soft contribution is 
given by the integral over the unshaded phase space in Fig.~\ref{softplus} with a minus sign. Since the unshaded region never touches the IR 
region, the soft contribution contains only UV divergence.

\subsubsection{The case $\alpha=1$}

The calculation procedure is the same for $\alpha=1$, hence the result can be extended to the case $\alpha=1$. The soft part at $\alpha=1$ 
is given as
\begin{equation}
M_{\mathrm{soft}}^{\alpha=1} =\frac{\alpha_s C_F}{2\pi} \Bigl(- \frac{2}{\euv^2} -\frac{2}{\euv} \ln \frac{4\mu^2}{y_c^2 Q^2} 
-\ln^2 \frac{4\mu^2}{y_c^2 Q^2} +\frac{\pi^2}{6}\Bigr),
\end{equation}
while the soft part in the JADE algorithm is given by
\begin{equation}
M_{\mathrm{soft}}^{\mathrm{JADE}} =\frac{\alpha_s C_F}{2\pi} \Bigl(- \frac{2}{\euv^2} -\frac{2}{\euv} \ln \frac{\mu^2}{j^2 Q^2} 
-\ln^2 \frac{\mu^2}{j^2 Q^2} +\frac{\pi^2}{6}\Bigr),
\end{equation}
which is identical to the soft part with $\alpha=1$ by replacing $j=y_c/2$. This should be true because the jet algorithm for $\alpha=1$ in 
the soft part is  identical to the JADE algorithm with $j=y_c/2$.
The soft functions in these two algorithms at next-to-leading order are given as
\begin{eqnarray}
\mathcal{S}^{(1)}_{\alpha =1} (Q, y_c,\mu)&=& \frac{\alpha_s C_F}{2\pi} \Bigl( 
-\ln^2 \frac{4\mu^2}{y_c^2 Q^2} +\frac{\pi^2}{6}\Bigr), \nonumber \\
\mathcal{S}^{(1)}_{\mathrm{JADE}} (Q,j,\mu) &=&\frac{\alpha_s C_F}{2\pi} \Bigl( 
-\ln^2 \frac{\mu^2}{j^2 Q^2} +\frac{\pi^2}{6}\Bigr).
\end{eqnarray}

\subsubsection{The case $1<\alpha<2$}
The contribution from the real gluon emission is given as
\begin{equation}
S_b =  2\frac{\alpha_s C_F}{2\pi}  \frac{e^{\gamma_{\mathrm{E}}\eps}}{\Gamma (1-\eps)}  \int_0^A dx x^{-1-\eps} \int_x^{g(x)} dy y^{-1-\eps}.
\end{equation}
It turns out that the result is the same as that in the case with $0<\alpha \le 1$. 

\subsubsection{The case $\alpha>2$}
The left-hand-side plot in Fig.~\ref{softplus} (e), $l_+$ approaches infinity as $l_-$ goes to zero. Therefore we may expect that there are both UV and
IR divergences. However, the plot of $\mathbf{l}_{\perp}^2$ with respect to $l_-$ in the right-hand side shows that
$\mathbf{l}_{\perp}^2$ can never reach infinity, hence the real gluon emission involves only the 
IR divergence. In order to handle the divergence correctly, we write $S_b$ as
\begin{eqnarray} \label{sbg2}
S_b &=&  2\frac{\alpha_s C_F}{2\pi}  \frac{e^{\gamma_{\mathrm{E}}\eps}}{\Gamma (1-\eps)}  \int_0^A dx x^{-1-\eps} \int_x^{g(x)} dy y^{-1-\eps} 
\nonumber \\
&=& 2\frac{\alpha_s C_F}{2\pi}  \frac{e^{\gamma_{\mathrm{E}}\eps}}{\Gamma (1-\eps)}  \int_0^A dx x^{-1-\eps} 
\Bigl[ \int_0^{g(x)} dy y^{-1-\eps}-\int_0^x dy y^{-1-\eps}\Bigr] \nonumber \\
&=& \frac{\alpha_s C_F}{2\pi} \Bigl[ \frac{1}{\alpha-2} \Bigl( \frac{2}{\eir^2} -\frac{2}{\eir} \ln a +  \ln^2 a -\frac{\pi^2}{6}\Bigr)
+\frac{1}{\eir^2} -\frac{1}{\alpha}\ln^2 a + \frac{\pi^2}{2} \Bigl( \frac{1}{3\alpha} -\frac{1}{2}\Bigr) \Bigr].
\end{eqnarray}
The amplitude for the soft function is given by
\begin{eqnarray}
M_{\mathrm{soft}}^{\alpha>2} &=&  2(S_a + S_b) \nonumber \\
&=&\frac{\alpha_s C_F}{2\pi}  \Bigl[ \frac{1}{\alpha-2} \Bigl( \frac{4}{\eir^2} -\frac{4}{\eir} \ln a 
+ 2\ln^2  a   -\frac{\pi^2}{3}\Bigr) - \frac{2}{\euv^2} +\frac{4}{\euv\eir}  \nonumber \\
&& -\frac{2}{\alpha} \ln^2 a+\pi^2 \Bigl( \frac{1}{3\alpha}-\frac{1}{2} \Bigr) \Bigr]. 
\end{eqnarray}

Note the difference of the soft amplitudes between the cases with $1<\alpha <2$ and $\alpha>2$. If we naively put $\euv =\eir =\eps$, the result is 
the same. However, if we distinguish the UV and IR divergences, the meaning of the results is totally different. The case with $1<\alpha <2$
contains only the UV poles, while the case with $\alpha>2$ is not IR finite. Therefore, when  $\alpha>2$, the soft function is not physically meaningful.
However, if we add the soft amplitude with the corresponding collinear amplitude, Eq.~(\ref{co2l}), the IR divergence cancels out. It means
that the dijet cross section can be computed with IR finiteness, but the factorization breaks down for $\alpha>2$.

One more comment is that the results are also divergent at $\alpha=2$. The soft function for $\alpha=2$ is not well defined in dimensional 
regularization. However, if we add the soft amplitude with the corresponding collinear amplitude, the divergent term in $\alpha$, proportional to
$1/(\alpha-2)$ also cancels out.  This will be discussed in detail in Section~\ref{alpha2}.

\section{Generalized $k_T$ jet algorithm with $\alpha \le 0$\label{negalp}}
The generalized $k_T$ algorithm for $\alpha \le 0$ is similar to the generic cone type jet algorithm or the Sterman-Weinberg jet algorithm 
as far as the shape
of the phase space is concerned \cite{Chay:2015ila}. The phase space for the naive collinear contribution, the zero-bin contribution and the soft
function is shown in Fig.~\ref{cominus}. We can extend the jet algorithm to $\alpha=0$ since the phase space for $\alpha=0$ is obtained by 
taking the limit $\alpha\rightarrow 0$ from the case with $\alpha <0$ because a jet veto is needed.  The jet algorithm with $\alpha=0$
is known as the Cambridge/Aachen jet algorithm.

\subsection{Jet function}
 
The naive collinear contribution from Fig.~\ref{intjet} (b) is given as
\begin{eqnarray}
\tilde{M}_b &=& \frac{\alpha_s C_F}{2\pi} \frac{e^{\gamma_{\mathrm{E}}\eps}}{\Gamma (1-\eps)}  \Bigl( 
\frac{\mu}{Q}\Bigr)^{\eps} 
\Bigl[ \int_0^{1/2} dx x^{-1-\eps} (1-x) \int_0^{f_2 (x)} dy y^{-1-\eps}  \nonumber \\
&&+\int_{1/2}^1 dx  x^{-1-\eps} (1-x) \int_0^{f_1 (x)} dy y^{-1-\eps} \Bigr]  \nonumber \\
&=& \frac{\alpha_s C_F}{2\pi} \Bigl[ \frac{1}{2\eir^2} +\frac{1}{\eir} \Bigl( 1 + \frac{1}{2}  \ln 
\frac{\mu^2}{2^{\alpha -2} y_c Q^2} \Bigr) +  \ln \frac{\mu^2}{2^{\alpha -2} y_c Q^2} 
+\frac{1}{4}  \ln^2 \frac{\mu^2}{2^{\alpha -2} y_c Q^2} \nonumber \\
&&+2 +2\ln 2 -\frac{5\pi^2}{24} +(\alpha-2) \Bigl(-1+\ln 2 +\frac{\pi^2}{12}\Bigr)\Bigr].
\end{eqnarray}

For the zero-bin contribution,   the boundary for the phase space is obtained by taking the zero-bin limit of 
$f_2 (x)$, which is given by
\begin{equation}
f_2^0 (x) = 2^{\alpha-2} y_c x.
\end{equation}
The zero-bin contribution is given as
\begin{eqnarray}
M_b^0 &=& \frac{\alpha_s C_F}{2\pi} \frac{e^{\gamma_{\mathrm{E}}\eps}}{\Gamma (1-\eps)}  \Bigl( 
\frac{\mu}{Q}\Bigr)^{\eps} 
\int_0^{\infty} dxx^{-1-\eps} \int_0^{f_2^0 (x)} dy y^{-1-\eps} \nonumber \\
&=& \frac{\alpha_s C_F}{2\pi} \frac{e^{\gamma_{\mathrm{E}}}}{\Gamma (1-\eps)}  \Bigl( 
\frac{\mu}{Q}\Bigr)^{\eps} 
\Bigl[ \int_0^{\eta} dx x^{-1-\eps} \int_0^{f_2^0 (x)} dy y^{-1-\eps}  \nonumber \\
&&+\int_{\eta}^{\infty} dx x^{-1-\eps}\Bigl( \frac{1}{\euv} -\frac{1}{\eir}
-\int_{f_2^0 (x)}^{\infty} dy y^{-1-\eps} \Bigr)\Bigr] \nonumber \\
&=& \frac{\alpha_s C_F}{2\pi}  \Bigl[ \frac{1}{2}\Bigl( \frac{1}{\euv} -\frac{1}{\eir}\Bigr)^2 +\frac{1}{2} 
\Bigl( \frac{1}{\euv} -\frac{1}{\eir}\Bigr) \ln 2^{\alpha-2} y_c\Bigr].
\end{eqnarray}

 From the real gluon emission in Fig.~\ref{intjet} (c), the naive collinear contribution is given by
\begin{eqnarray}
\tilde{M}_c &=& \frac{\alpha_s C_F}{2\pi} \frac{e^{\gamma_{\mathrm{E}}\eps}}{\Gamma (1-\eps)}  \Bigl(\frac{\mu}{Q}\Bigr)^{\eps} (1-\eps)
\Bigl[ \int_0^{1/2} dx x^{1-\eps} \int_0^{f_2 (x)} dy y^{-1-\eps}+ \int_{1/2}^1 dx x^{1-\eps} \int_0^{f_1 (x)} dy y^{-1-\eps}\Bigr]\nonumber \\
&=& \frac{\alpha_s C_F}{2\pi}\Bigl( -\frac{1}{2\eir} -\frac{1}{2} \ln \frac{\mu^2}{2^{\alpha-2} y_c Q^2}+\frac{\alpha -3}{2} 
-\frac{\alpha}{2}\ln 2\Bigr).
\end{eqnarray}
Compared to the corresponding calculation for $\alpha>0$ in Eq.~(\ref{mcal+}), the only difference is in the coefficient of $\ln 2$.

Combining the wave function renormalization and the residue, the collinear amplitude for $\alpha \le 0$ is given as
\begin{eqnarray}
M_{\mathrm{coll}}^{\alpha \le 0} &=& \frac{\alpha_s C_F}{2\pi} \Bigl[ \frac{1}{\euv^2} +\frac{1}{\euv} \Bigl( \frac{3}{2}  
 + \ln \frac{\mu^2}{2^{\alpha -2} y_c Q^2} \Bigr) 
+\frac{3}{2} \ln \frac{\mu^2}{2^{\alpha -2} y_c Q^2}+\frac{1}{2} \ln^2 \frac{\mu^2}{2^{\alpha -2} y_c Q^2}\nonumber \\
&& + \frac{13-3\alpha}{2}  +\frac{3\alpha}{2} \ln 2 +\frac{\pi^2}{2}\Bigl( \frac{\alpha}{3}-\frac{3}{2}\Bigr)\Bigr].
\end{eqnarray}
And the jet function at one loop for $\alpha \le 0$ is given as
\begin{eqnarray}
\mathcal{J}^{(1)} (Q,y_c,\mu) &=& \frac{\alpha_s C_F}{2\pi} \Bigl[  
\frac{3}{2} \ln \frac{\mu^2}{2^{\alpha -2} y_c Q^2}+\frac{1}{2} \ln^2 \frac{\mu^2}{2^{\alpha -2} y_c Q^2} \nonumber \\
&&  + \frac{13-3\alpha}{2}  +\frac{3\alpha}{2} \ln 2 +\frac{\pi^2}{2}\Bigl( \frac{\alpha}{3}-\frac{3}{2}\Bigr)\Bigr].
\end{eqnarray}

\subsection{Soft function}

The contribution from the real gluon emission is given by
\begin{eqnarray}
S_b &=& \frac{\alpha_s C_F}{2\pi}  \Bigl[\Bigl(\frac{1}{\euv} -\frac{1}{\eir} \Bigr)^2 +\frac{1}{\euv} \ln (2^{\alpha-2} y_c )  \nonumber \\
&&+2 \ln \frac{\mu}{2\beta Q} \ln (2^{\alpha-2} y_c )  -\frac{1}{2} \ln^2 (2^{\alpha-2} y_c )  -\frac{\pi^2}{6}\Bigr].
\end{eqnarray}
With the virtual correction, the amplitude for the soft function is written as
\begin{equation}
M_{\mathrm{soft}}^{\alpha \le 0} = \frac{\alpha_s C_F}{2\pi}  \Bigl( \frac{2}{\euv} \ln (2^{\alpha-2} y_c) 
+ 4\ln \frac{\mu}{2\beta Q} \ln (2^{\alpha-2} y_c)  -
\ln^2 (2^{\alpha-2} y_c) -\frac{\pi^2}{3}\Bigr),
\end{equation}
and the soft function at one loop for $\alpha \le 0$ is given by
\begin{equation}
\mathcal{S}^{(1)} (Q, y_c, \beta, \mu) = \frac{\alpha_s C_F}{2\pi}  \Bigl(  4\ln \frac{\mu}{2\beta Q} \ln (2^{\alpha-2} y_c)  -
\ln^2 (2^{\alpha-2} y_c) -\frac{\pi^2}{3}\Bigr).
\end{equation}
In the limit $\alpha \rightarrow 0$, which corresponds to the Cambridge/Aachen jet algorithm, the soft function is given as
\begin{equation}
\mathcal{S}^{(1)}_{\alpha=0} (Q, y_c, \beta, \mu) 
= \frac{\alpha_s C_F}{2\pi}  \Bigl(   4\ln \frac{\mu}{2\beta Q} \ln \frac{y_c}{4}   -
\ln^2 \frac{y_c}{4}  -\frac{\pi^2}{3}\Bigr). 
\end{equation}
Note that the limit $\alpha\rightarrow 0$ should be taken from $\alpha <0$. For $\alpha>0$, there is a pole in $1/\alpha$, and it may be 
dangerous to take the limit $\alpha \rightarrow 0$. However, when the limit $\alpha \rightarrow 0$ is taken, it coincides with the phase 
space for $\alpha <0$ and the jet veto is needed. Therefore the soft function at $\alpha=0$ should be taken from the soft part with $\alpha<0$.
 
\section{Phase space and structure of divergence \label{psdiv}}
The structure of divergence can be inferred from the phase space argument \cite{Hornig:2009kv,Hornig:2009vb}. Since the phase spaces for
different $\alpha$ yield different shapes of the phase space, we will discuss and compare the cases for various values of $\alpha$. 

Let us first consider the phase space for the naive collinear contribution in Fig.~\ref{phplus}. In the first column, the phase space is 
plotted in $(l_-, l_+)$ space, and the second column in $(l_-, \mathbf{l}_{\perp}^2)$ space.  When we compute the naive collinear contribution, 
there is an integration with respect to the gluon momentum $l$. If we first integrate over $\mathbf{l}_{\perp}$, which takes care of the 
delta function in the integrand, the remaining integral is to be 
performed with respect to $l_-$ and $l_+$,  where the phase space in the first column is relevant. If we first integrate over $l_+$, the 
remaining integral is to be performed with respect to $l_-$ and $\mathbf{l}_{\perp}$, where the phase space in the second column is appropriate.

In  $(l_-, l_+)$ space, the UV (IR) divergence comes from the region where $l_-$ and $l_+$ approaches infinity (zero). If one of them approaches
infinity, while the other approaches zero, it is possible that there is a mixed UV-IR divergence. We can infer the structure of divergence  in
 $(l_-, l_+)$ space from the first column of Fig.~\ref{phplus}.  For $\alpha \le 1$, it is clear that, if there is divergence, it should be of the IR origin
 from Fig.~\ref{phplus} (a) and (b) since $l_-$ and $l_+$ can never reach infinity.  It is explicitly verified 
in Eqs.~(\ref{naimb}), (\ref{mcal+}). This is also 
evident from the second column of the corresponding phase spaces since $l_-$ and $\mathbf{l}_{\perp}^2$ never reach infinity. 
With the same argument, when $\alpha>2$ in Fig.~\ref{phplus} (e), both plots imply that there are UV, IR and mixed divergences. 

However, the two plots lead to seemingly inconsistent results in Fig.~\ref{phplus} (c), which corresponds to $1<\alpha < 2$. 
In  $(l_-, l_+)$ space, $l_+$ can reach infinity while $l_-$ cannot. Therefore we naively expect that there might be IR and mixed divergences.
On the other hand, in $(l_-, \mathbf{l}_{\perp}^2)$ space, $\mathbf{l}_{\perp}^2$ never reaches infinity and there should be only IR divergence.
In this case, we have to rely on the phase space in $(l_-, \mathbf{l}_{\perp}^2)$ space because there is an additional relation 
$l_+ l_- = \mathbf{l}_{\perp}^2$ for the real gluon emission.

From the phase space constraint in Eq.~(\ref{cocon}), it reads for $l_- <Q/2$
\begin{equation}
l_+ < \frac{Q^{\alpha}}{2^{2-\alpha}} y_c \Bigl( 1-\frac{l_-}{Q}\Bigr)^2 l_-^{1-\alpha},
\end{equation}
and indeed $l_+$ can approach infinity as $l_-$ approaches zero for $1<\alpha <2$. However, $\mathbf{l}_{\perp}^2$ is bounded by
\begin{equation}
\mathbf{l}_{\perp}^2 = l_+ l_- < \frac{Q^{\alpha}}{2^{2-\alpha}} y_c \Bigl( 1-\frac{l_-}{Q}\Bigr)^2 l_-^{2-\alpha},
\end{equation}
which cannot reach infinity as $l_-$ approaches zero.  Since the divergence behavior is determined by the behavior in $\mathbf{l}_{\perp}^2$
in $D-2$ dimensions, and $\mathbf{l}_{\perp}^2=l_+ l_-$, the inference on the divergence behavior is more reliable in 
$(l_-, \mathbf{l}_{\perp}^2)$ space rather than in  $(l_-, l_+)$ space.  This can also be confirmed in Ref.~\cite{Chay:2012mh}, in which 
the jet function has been computed with the transverse momentum fixed. Since the transverse momentum is fixed at a finite value, 
there is no UV divergence, but IR divergence.
The computation was performed for the inclusive jet function with the transverse momentum fixed, hence there 
is no constraint on the phase space as in the jet algorithm. However, the result supports that there is only IR divergence 
when $\mathbf{l}_{\perp}^2$ never reaches infinity.  As can be seen in Fig.~\ref{phplus} (a) to (c), the
phase spaces for $1<\alpha <2$ in $(l_-, \mathbf{l}_{\perp}^2)$ space shows similar shapes and there is only IR divergence. 

For the phase space of the zero-bin contribution in Fig.~\ref{zeroplus}, all the plots show that the zero-bin contribution should contain
both UV and IR divergences. 
For the soft contribution shown in Fig.~\ref{softplus}, we can conclude that there are both UV and IR divergences for $0<\alpha<2$,
while there is only IR divergence for $\alpha \ge 2$, as can be seen in Eq.~(\ref{sbg2}).  
However, we can apply the phase space method as in Ref.~\cite{Hornig:2009kv}. 
The virtual correction for the soft part has the same integral as in the real gluon emission except the minus sign, and it covers the entire phase space.
Therefore the soft part is obtained by integrating the matrix element over the unshaded phase space in Fig.~\ref{softplus} with a minus sign.
Then the soft part for $0<\alpha <2$ has only UV divergence since the unshaded region never reaches zero. It means that the cancellation
of the IR divergence occurs between the virtual and real contributions. On the other hand, for $\alpha \ge 2$, we expect that there is not only
UV divergence, but also IR divergence, which breaks the factorization.

The case with $\alpha=2$ (the $k_T$ algorithm) is tricky because it contains the pole in $1/(\alpha-2)$, and it cannot be computed in
dimensional regularization.  However, from the phase space diagrams, we expect that the naive collinear contribution  contains only IR divergence,
while the soft contribution contains both UV and IR divergences.  This will be discussed in detail in the next section.

\section{Comment on the $k_T$ algorithm with $\alpha=2$\label{alpha2}}
We have considered the jet and the soft functions in the generalized jet algorithm by varying $\alpha$. It turns out that the jet algorithm 
for $\alpha <2$ respects
the factorization, that is, the jet and the soft functions are IR finite, and it breaks the factorization for $\alpha \ge 2$ since each part contains
IR divergence. The case with $\alpha=2$ shows a peculiar behavior and it is worth mentioning a few characteristics. 

At $\alpha=2$, in the $k_T$ algorithm, there is an uncontrollable integral of the form
\begin{equation}
\int_0 \frac{dx}{x}
\end{equation}
in the naive collinear, the zero-bin and the soft contributions, even though the spacetime dimension is extended to $D=4-2\eps$. It is also noted
in Ref.~\cite{Cheung:2009sg}. 

If we use the offshellness to regulate the IR divergence in the soft Wilson line by replacing the propagator, the soft part at $\alpha=2$ is given by 
 \begin{equation}
M_{\mathrm{soft}}^{\alpha=2} =\frac{\alpha_s C_F}{2\pi} \Bigl( -\frac{2}{\euv^2} -\frac{2}{\euv} \ln \frac{\mu^2}{\Delta_1 \Delta_2} -
\ln^2 \frac{\mu^2}{\Delta_1 \Delta_2} +\frac{1}{2}\ln^2 \frac{y_c Q^2}{\Delta_1^2}+\frac{1}{2}\ln^2 \frac{y_c Q^2}{\Delta_2^2}
+\cdots\Bigr),
\end{equation}
where $\Delta_i = p_i^2/Q$ represent the offshellness of the quark and the antiquark. And this is consistent with the result in 
Ref.~\cite{Cheung:2009sg}.
The result clearly shows that the soft part contains the UV and IR divergences.  
If we look at the plots for the phase space in Figs.~\ref{phplus} (d), \ref{zeroplus} (d) and \ref{softplus} (d), it can be inferred that
the naive collinear contribution contains only IR divergence, the zero-bin and the soft contribution have both UV and IR divergences.

In spite of the possible argument on the existence of the UV and IR divergence in the jet and the soft functions for $\alpha=2$, we interpret
the result differently in the sense  that they are not well defined for $\alpha=2$.
If we consider the jet and soft functions as a function of $\alpha$ in the generalized $k_T$ algorithm, they are not continuous at $\alpha=2$
in the sense that
\begin{equation}
\lim_{\alpha\rightarrow 2^-} (M_{\mathrm{coll}}^{\alpha <2}, M_{\mathrm{soft}}^{\alpha<2} ) \neq \lim_{\alpha\rightarrow 2^+}  
 (M_{\mathrm{coll}}^{\alpha >2}, M_{\mathrm{soft}}^{\alpha>2} ).
\end{equation}
When we look at the plots for the phase space, for example, Fig.~\ref{phplus}, the phase space allows finite $\mathbf{l}_{\perp}^2$ only for
$\alpha<2$, but $\mathbf{l}_{\perp}^2$ can reach infinity for $\alpha>2$. Therefore the divergence structure abruptly changes at $\alpha=2$. 
This discontinuity appears as a pole in $\alpha-2$ in our treatment of the generalized jet algorithm as a function of $\alpha$.  In contrast,
the behavior of the jet function and the soft function near $\alpha=1$ is smooth. The shape of the phase space in the $(l_-, l_+)$ space  
changes near $\alpha=1$, that is, $l_+$ approaches a finite value instead of zero near $l_-=0$. However, in the $(l_-,\mathbf{l}_{\perp}^2)$ space,
there is no abrupt change in shape. Therefore the jet algorithm is continuous at 
$\alpha=1$, hence the factorization works in the JADE algorithm, too.

Interestingly enough, the pole in $\alpha-2$ as well as the
IR divergence cancels in the sum of the collinear part and the soft part to yield the jet cross section. Therefore
the jet cross section  is IR finite even at $\alpha =2$. We may conclude that the $k_T$ algorithm works
fine, but it is misleading.  The main point is that
 the factorization, which lies behind the jet algorithm, breaks down for $\alpha \ge 2$ including the $k_T$ algorithm.  

However, note that the claim that the $k_T$ algorithm breaks factorization may be too strong because the separation of the UV and IR 
divergence is spoiled by the existence of the pole in $\alpha-2$. In dimensional regularization, the extraction of the divergence is impossible
at $\alpha=2$, but there may be other ways to obtain the divergence unambiguously. For example,
in Ref.~\cite{Cheung:2012ag}, the authors have tried to regulate this integral with the rapidity renormalization
technique. It may turn out to be useful, but we will not pursue in this direction here, and our claim stands as it is as far as we use the
dimensional regularization to extract UV and IR divergences.

\section{Renormalization group behavior of  the dijet cross sections\label{xsec}}
Let us collect all the results for the resummation of the large logarithms of the form $\alpha_s^n \ln^k y_c$ with $n \le k \le 2n$ through
the renormalization group behavior. At one loop, the hard function is given by
\begin{equation}
H^{(1)} (Q^2, \mu) =  \frac{\alpha_s C_F}{2\pi} \Bigl(-\ln^2 \frac{\mu^2}{Q^2} -3 \ln \frac{\mu^2}{Q^2} -8 +\frac{7 \pi^2}{6}\Bigr).
\end{equation}
The jet function and the soft function for $0<\alpha <2$ at one loop are given as
\begin{eqnarray} \label{js+}
\mathcal{J}^{(1)}_n (Q,y_c,\mu) &=&  
\frac{\alpha_s C_F}{2\pi}  \Bigl[\frac{1}{\alpha-2} \Bigl( - \ln^2 \frac{\mu^2}{2^{\alpha-2} y_c Q^2}
 +\frac{\pi^2}{6}\Bigr) +  \frac{3}{2}  \ln \frac{\mu^2}{2^{\alpha-2} y_c Q^2} \nonumber \\
&&+\frac{13-3\alpha}{2}   -\frac{3\alpha}{2}\ln 2  +(\alpha-4)\frac{\pi^2}{6}  \Bigr], \nonumber \\
\mathcal{S}^{(1)} (Q,y_c,\mu)  &=& \frac{\alpha_s C_F}{2\pi} \Bigl[\frac{4}{\alpha(\alpha-2)}  
\ln^2  2^{\alpha-2} y_c \Bigl(\frac{Q}{\mu}\Bigr)^{\alpha}    
 -\pi^2 \Bigl( \frac{2}{3\alpha(\alpha-2)} +\frac{1}{2} \Bigr) \Bigr]. 
\end{eqnarray}
The jet function $\mathcal{J}_{\bar{n}}^{(1)}$ in the $\overline{n}$ direction is the same as $\mathcal{J}^{(1)}_n$.
The dijet cross section in Eq.~(\ref{facjet}) at next-to-leading order (NLO) for $0 <\alpha <2$ is written as
\begin{equation}
\sigma^{(1)}_{0<\alpha<2}  
= \sigma_0 \frac{\alpha_s C_F}{2\pi} \Bigl[ -\frac{2}{\alpha}\ln^2 (2^{\alpha-2} y_c)  -3 \ln (2^{\alpha-2} y_c)
+5-3\alpha -3\alpha \ln 2   + \frac{(\alpha-1)^2}{3\alpha}\pi^2\Bigr]. 
\end{equation}
For example, the dijet cross section for $\alpha=1$, which is close to the JADE algorithm, is given by
\begin{equation}
\sigma_{\alpha=1}^{(1)} =\sigma_0 \frac{\alpha_s C_F}{2\pi} \Bigl( -2 \ln^2 \frac{y_c}{2} -3\ln \frac{y_c}{2} +2-3\ln 2  \Bigr),
\end{equation}
while the cross section exactly in the  JADE algorithm is given by
\begin{equation}
\sigma_{\mathrm{JADE}}^{(1)} = \sigma_0\frac{\alpha_s C_F}{2\pi} \Bigl( -2 \ln^2 j -3\ln j -1 +\frac{\pi^2}{3}\Bigr).
\end{equation}
This agrees with the full QCD calculation \cite{Kramer:1986sg}.

For $\alpha <0$, the jet function and the soft function at one loop are given as
\begin{eqnarray} \label{js-}
\mathcal{J}^{(1)} (Q,y_c,\mu) &=& \frac{\alpha_s C_F}{2\pi} \Bigl[  
\frac{3}{2} \ln \frac{\mu^2}{2^{\alpha -2} y_c Q^2}+\frac{1}{2} \ln^2 \frac{\mu^2}{2^{\alpha -2} y_c Q^2}   \nonumber \\
&&+ \frac{13-3\alpha}{2}  +\frac{3\alpha}{2} \ln 2 +\frac{\pi^2}{2} \Bigl(\frac{\alpha}{3} -\frac{3}{2}\Bigr)\Bigr], \nonumber \\
\mathcal{S}^{(1)} (Q, y_c, \beta, \mu) &=& \frac{\alpha_s C_F}{2\pi}  \Bigl(  4\ln \frac{\mu}{2\beta Q} \ln (2^{\alpha-2} y_c)  -
\ln^2 (2^{\alpha-2} y_c) -\frac{\pi^2}{3}\Bigr).
\end{eqnarray}

The dijet cross section at NLO for $\alpha \le 0$ is given as
\begin{equation}
\sigma^{(1)}_{\alpha \le 0} 
= \sigma_0\frac{\alpha_s C_F}{2\pi} \Bigl[\Bigl( -3- 4\ln 2\beta \Bigr)\ln (2^{\alpha-2} y_c)   -5-3\alpha 
+3\alpha \ln 2+\frac{\pi^2}{3} (\alpha-2)\Bigr]. 
\end{equation}
An interesting case is the result of the anti-$k_T$ jet algorithm with $\alpha=-2$. The dijet cross section in the anti-$k_T$ algorithm at order $\alpha_s$ is explicitly given by
\begin{equation}
\sigma^{(1)}_{\mathrm{anti-}k_T} = \sigma_0 \frac{\alpha_s C_F}{2\pi} \Bigl[( -3 -4 \ln 2 \beta)  \ln \frac{y_c}{16} 
+11-6\ln 2 -\frac{4\pi^2}{3}\Bigr].
\end{equation} 
Also the Cambridge/Aachen jet algorithm corresponds to the case $\alpha=0$, and the dijet cross section in this jet algorithm is given as
\begin{equation}
\sigma^{(1)}_{\mathrm{C/A}} = \sigma_0\frac{\alpha_s C_F}{2\pi} \Bigl[( -3  -4 \ln 2 \beta) \ln \frac{y_c}{4} +5 -\frac{2\pi^2}{3}\Bigr].
\end{equation} 

For $\alpha  \ge 2$, the factorization breaks down since the jet and the soft functions contain IR divergence. 
However, when the
jet function and the soft function are added, these divergent terms cancel and the finite dijet cross section is obtained.  
The sum of the collinear and the soft parts is given as
\begin{eqnarray}
2M_{\mathrm{coll}}^{\alpha>2} + M_{\mathrm{soft}}^{\alpha>2} &= &\frac{\alpha_s C_F}{2\pi} \Bigl[ \frac{2}{\euv^2} +\frac{2}{\euv} \Bigl( 
\ln \frac{\mu^2}{Q^2} +\frac{3}{2} \Bigr)+\ln^2 \frac{\mu^2}{Q^2} +3 \ln \frac{\mu^2}{Q^2}  -\frac{2}{\alpha}\ln^2 2^{\alpha-2} y_c 
\nonumber \\
&&-3 \ln 2^{\alpha-2} y_c +13 -3\alpha -3\alpha \ln 2 +\frac{\pi^2}{3}\Bigl(\alpha +\frac{1}{\alpha} -\frac{11}{2}\Bigr) \Bigr].
\end{eqnarray}
Note that only this combination is free of IR divergence as well as of the pole in $\alpha -2$. Therefore the dijet cross section has a finite limit
as $\alpha \rightarrow 2$. The dijet cross section for $\alpha > 2$ is given as
\begin{eqnarray}
\sigma^{(1)}_{\alpha > 2} &=& \sigma_0\frac{\alpha_s C_F}{2\pi}  \Bigl[ -\frac{2}{\alpha}\ln^2 2^{\alpha-2} y_c
-3 \ln 2^{\alpha-2} y_c  +5-3\alpha (1+ \ln 2) + \frac{(\alpha-1)^2}{3\alpha} \pi^2  \Bigr].
\end{eqnarray}
Note that the dijet cross section is the same for $\alpha<2$ and $\alpha>2$, and it is continuous at $\alpha=2$.
Therefore,  for the $k_T$ algorithm, the dijet cross section at order $\alpha_s$ is given as
\begin{equation}
\sigma^{(1)}_{k_T} = \sigma_0\frac{\alpha_s C_F}{2\pi} \Bigl( -\ln^2 y_c -3\ln y_c  -1-6\ln 2 +\frac{\pi^2}{6}\Bigr).
\end{equation}

The fixed-order dijet cross sections for various values of $\alpha$ are independent of the renormalization scale $\mu$. However, they contain
logarithms of $y_c$. When $y_c \sim \mathcal{O} (\lambda^2)$ becomes small, the fixed-order results are not to be trusted and the 
resummation of the large logarithms should be performed. This is achieved by solving the renormalization group (RG) equations for the 
hard, jet and soft functions. We present the solution of the RG equation and show the numerical results of the resummation effect with the
theoretical uncertainties at next-to-leading logarithm (NLL) order.

The RG equation for the hard, jet and soft functions is of the form
\begin{equation}
\frac{d}{d\ln \mu} f_i (\omega_i,\mu) = \Bigl[ a_i (\alpha_s, \alpha) \ln \frac{\omega_i^2}{\mu^2} + b_i (\alpha_s, \alpha) 
\Bigr] f_i(\omega_i,\mu),
\end{equation}
where $f_i (\omega_i,\mu)$ is the corresponding function and $\omega_i = Q, \mu_J, \mu_S$ respectively for $i=H, J, S$.  
The solution of the RG equation can be expressed in terms of 
the following quantities which are defined as \cite{Becher:2006mr}
\begin{eqnarray}
\frac{d}{d\ln \mu} S_i(\nu, \mu) &=& -a_i \bigl( \alpha_s(\mu) \bigr) \ln \frac{\mu}{\nu},  \nonumber \\
\frac{d}{d\ln \mu} A_i(\nu, \mu) &=&  -a_i \bigl( \alpha_s(\mu) \bigr), \ \
\frac{d}{d\ln \mu} B_i(\nu, \mu) =  -b_i \bigl( \alpha_s(\mu) \bigr).
\end{eqnarray}
Then $f_i (\omega_i,\mu)$ is written as
\begin{equation} \label{rgf}
f_i (\omega_i,\mu) = \exp \Bigl[ 2 S_i(\mu_i, \mu) -B_i(\mu_i, \mu) \Bigr] \Bigl(\frac{\omega_i^2}{\mu_i^2}\Bigr)^{-A_i(\mu_i, \mu)} 
f_i (\omega_i,\mu_i),
\end{equation}
where the scales $\mu_i$ for $i=H, J, S$ are the  hard, jet and  soft scales respectively

The $\beta$ function and the anomalous dimensions can be expanded as a series in $\alpha_s$ as
\begin{eqnarray}
\beta(\alpha_s) &=& \frac{d}{d\ln \mu} \alpha_s = -2 \alpha_s \Bigl[ \beta_0 \frac{\alpha_s}{4\pi} 
+ \beta_1 \Bigl(\frac{\alpha_s}{4\pi}\Bigr)^2 +\cdots \Bigr], \nonumber \\
a_i (\alpha_s) &=& a_i^0  \frac{\alpha_s}{4\pi} + a_i^1 \Bigl(\frac{\alpha_s}{4\pi}\Bigr)^2 +\cdots,  \ \ 
b_i (\alpha_s) = b_i^0  \frac{\alpha_s}{4\pi} +b_i^1 \Bigl(\frac{\alpha_s}{4\pi}\Bigr)^2 +\cdots,
\end{eqnarray}

The anomalous dimension for the hard coefficient at order $\alpha_s$  is given by
\begin{equation}
\gamma_H = \frac{\alpha_s C_F}{2\pi} \Bigl( 4\ln \frac{Q^2}{\mu^2}-6\Bigr).
\end{equation}
The anomalous dimensions of the jet and the soft functions for $0 < \alpha <2$ at order $\alpha_s$ are given as
\begin{eqnarray} \label{ajpos}
\gamma_J &=& \frac{\alpha_s C_F}{2\pi} \Bigl( \frac{4}{\alpha-2} \ln \frac{\mu_J^2}{\mu^2} +3 \Bigr), \nonumber \\
\gamma_S &=& -\frac{\alpha_s C_F}{2\pi} \frac{4\alpha}{\alpha -2} \ln \frac{\mu_S^2}{\mu^2}, 
\end{eqnarray}
where the jet scale $\mu_J$ and the soft scale $\mu_S$ are defined as
\begin{equation}
\mu_J=\Bigl( 2^{\alpha-2} y_c\Bigr)^{1/2}  Q, \ \mu_S =\Bigl(2^{\alpha -2} y_c\Bigr)^{1/\alpha} Q.
\end{equation}
For $\alpha \le 0$, they are given as
\begin{eqnarray}
\gamma_J &=& \frac{\alpha_s C_F}{2\pi} \Bigl(  -2 \ln \frac{\mu_J^2}{\mu^2} +3 \Bigr), \nonumber \\
\gamma_S &=& 
\frac{\alpha_s C_F}{2\pi}   4\ln  \Bigl( 2^{\alpha-2} y_c\Bigr).
\end{eqnarray}
It can be explicitly verified that $\gamma_H + 2\gamma_J +\gamma_S =0$ for $\alpha <2$, and the dijet cross section is 
independent of the renormalization scale. However, the anomalous dimensions for the jet and the soft functions for $\alpha \ge 2$ are not defined
since the corresponding amplitudes include IR divergence. In this case the anomalous dimension of the combined collinear and soft parts is given by
\begin{equation}
\gamma_{2J+S} = \frac{\alpha_s C_F}{2\pi}  \Bigl( -4\ln \frac{Q^2}{\mu^2} +6\Bigr),
\end{equation}
which exactly cancels that of the hard function $H^{(1)}$.

The anomalous dimensions for $0<\alpha<2$ can be cast into the form
\begin{eqnarray}
\gamma_H (Q,\mu) &=& \Gamma_{\mathrm{cusp}} (\alpha_s) \ln \frac{Q^2}{\mu^2} + \Gamma^H (\alpha_s), \nonumber \\
\gamma_J (\mu_J, \mu) &=& \frac{1}{\alpha-2} \Gamma_{\mathrm{cusp}} (\alpha_s) \ln \frac{\mu_J^2}{\mu^2} + 
\Gamma^J (\alpha_s), \nonumber \\
\gamma_S (\mu_S, \mu) &=& -\frac{\alpha}{\alpha-2}\Gamma_{\mathrm{cusp}} (\alpha_s) \ln \frac{\mu_S^2}{\mu^2} 
+ \Gamma^S (\alpha_s),
\end{eqnarray}
and for $\alpha\le 0$,
\begin{eqnarray}
\gamma_J (\mu_J, \mu) &=&  -\frac{1}{2}\Gamma_{\mathrm{cusp}} (\alpha_s) \ln \frac{\mu_J^2}{\mu^2} + 
\Gamma^J (\alpha_s), \ 
\gamma_S (\mu_S, \mu) =  \Gamma^S (\alpha_s),
\end{eqnarray}
where $\Gamma_{\mathrm{cusp}}(\alpha_s)$ is the cusp anomalous dimension of the hard function $H(Q,\mu)$. This form is obtained
since $\gamma_H + 2\gamma_J +\gamma_S =0$, and since the cusp anomalous dimensions are independent of $\alpha$, that is, the jet algorithm.
To NLL order with large logarithms, we need the cusp anomalous dimension to two-loop order, the remaining anomalous dimensions to
one loop order and the hard, jet and soft functions at tree level \cite{Becher:2006mr}. The details for the numerical analysis are 
provided in Appendix.

 The hard, jet and soft scales can be inferred by requiring that the logarithmic terms in the fixed-order results are not significant. 
The hard scale is set by $\mu_H = Q$, and it is varied from $\mu_H/2$ to $2\mu_H$ to evaluate the theoretical uncertainty. The jet scale
is given by $\mu_J = (2^{\alpha-2} y_c)^{1/2}Q$ for $\alpha <2$, and is also varied from $\mu_J/2$ to $2\mu_J$. 

 \begin{figure}[t] 
\begin{center}
\includegraphics[width=16cm]{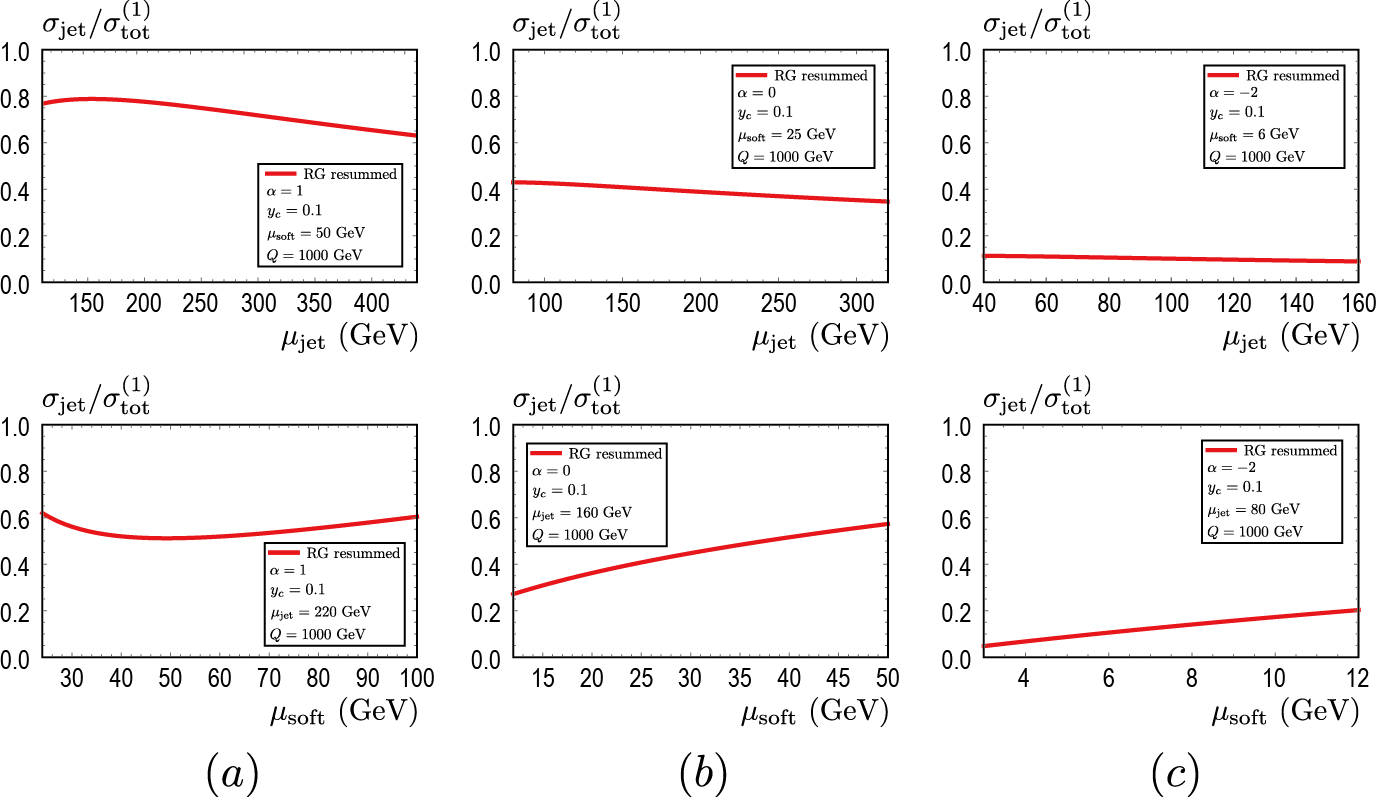}
\end{center}  
\vspace{-0.3cm}
\caption{\baselineskip 3.0ex The  dependence of the resummed dijet cross sections on the renormalization scales  for (a) $\alpha=0$ (b) 
$\alpha = -1$ (c) $\alpha=-2$. We set $Q=1000$ GeV, and $y_c=0.1$, and for the soft function $\beta =0.1$. \label{csecscale}}
\end{figure}

The choice of the soft scale is more delicate. For $0<\alpha <2$, from the argument of the logarithm  in the soft function in 
Eq.~(\ref{js+}), the soft scale is set as $\mu_S$.
From the power counting, we require that $\mu_J \sim \lambda Q$ and $\mu_S \sim \lambda^2 Q$, where $\lambda$ is a small
parameter in SCET. For $\alpha<0$, the soft scale is set by $\mu_S = 2 \beta Q$. As discussed in the jet algorithm,  $\beta$ is also of order 
$\lambda^2$.  Since $\lambda^2$ is set by the jet scale as $2^{\alpha-2} y_c$ , we require that $2\beta \sim 2^{\alpha-2} y_c$ in the numerical
analysis. We also vary the soft scale from $\mu_S/2$ to $2\mu_S$ to evaluate the theoretical uncertainty.

Fig.~\ref{csecscale} shows the theoretical uncertainty of the jet cross sections for three values of $\alpha$ at NLL as we vary the jet and soft
scales. The cross sections are normalized by dividing the jet cross section by the total cross section at order $\alpha_s$, which is given by
\begin{equation}
\sigma_{\mathrm{tot}}^{(1)} = \sigma_0 \Bigl(1+\frac{\alpha_s (Q)}{\pi}\Bigr).
\end{equation}
Therefore the ratio $\sigma_{\mathrm{jet}}/\sigma_{\mathrm{tot}}^{(1)} $ represents the fraction for the two jets.
The values of $Q$ and $y_c$ are chosen as 1000 GeV, and 0.1 respectively for illustrative purposes. The hard scale is fixed at $Q$.
And the cases with $\alpha=1$, 0 and $-2$ are shown, which correspond to the JADE, the Cambrige/Aachen, and the anti-$k_T$ algorithms 
respectively. 

The upper plots show the dependence on the jet scale $\mu_{\mathrm{jet}}$ as it varies from $\mu_J/2$ to $2\mu_J$ while the soft scale
$\mu_{\mathrm{soft}}$ is fixed at $\mu_S$. The lower plots show the dependence on the soft scale $\mu_{\mathrm{soft}}$ as it varies
from  $\mu_S/2$ to $2\mu_S$ while the jet scale $\mu_{\mathrm{jet}}$ is fixed at $\mu_J$. As can be seen in the plots, the dependence
on the jet scale is mild, but the cross section is rather sensitive to the choice of the soft scale for $\alpha=0$ and $-2$. 

\begin{figure}[b] 
\begin{center}
\includegraphics[width=16cm]{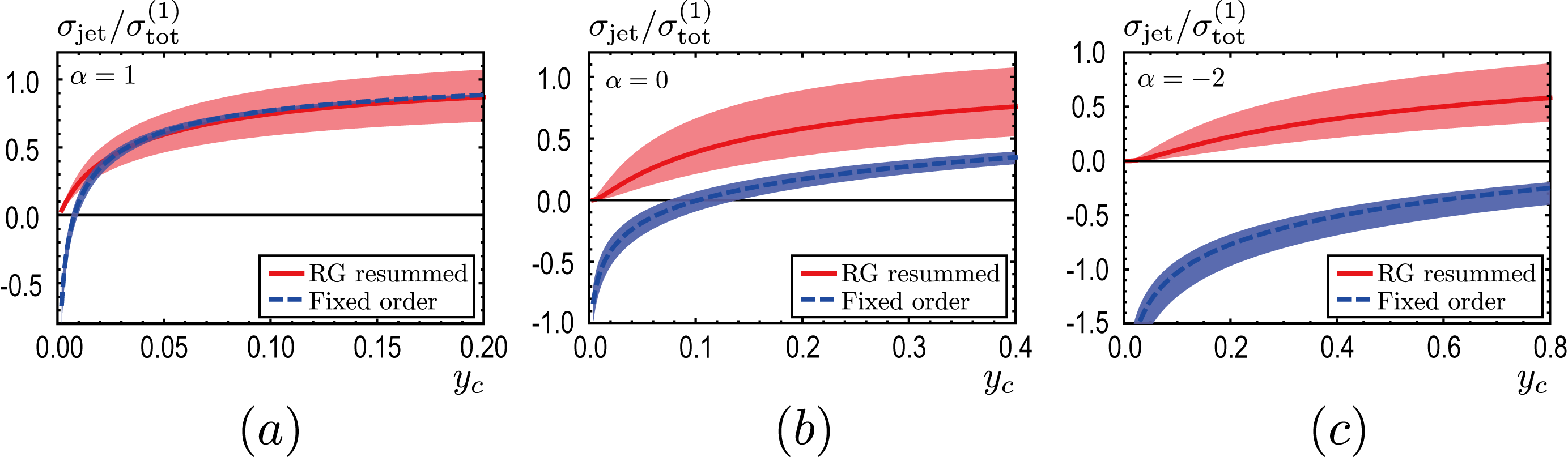}
\end{center}  
\vspace{-0.3cm}
\caption{\baselineskip 3.0ex The dijet cross sections with (a) $\alpha=1$ (b) $\alpha=0$ (c) $\alpha=-2$ corresponding to Fig.~\ref{csecscale}.  
The bands show the theoretical uncertainties. The solid line represents the resummed cross section with $\mu_{\mathrm{hard}}=Q$, 
$\mu_{\mathrm{jet}} = \mu_J$, $\mu_{\mathrm{soft}} = \mu_S$. The dashed line is the fixed-order cross section with $\mu=Q$.   
\label{csec}}
\end{figure}

Fig.~\ref{csec} shows the theoretical uncertainty of the resummed and the fixed-order dijet cross sections. In resummed results, the hard,
jet and soft scales are varied as $Q/2 < \mu_{\mathrm{hard}} <2Q$, $\mu_J/2 <\mu_{\mathrm{jet}} <2\mu_J$ and $\mu_S/2 <
\mu_{\mathrm{soft}} <2\mu_S$ respectively. In the fixed-order result, we set $\mu_{\mathrm{hard}} =
\mu_{\mathrm{jet}} =\mu_{\mathrm{soft}}$ and $\mu_{\mathrm{hard}}$ is varied from $Q/2$ to $2Q$. The small parameter 
$\lambda$ corresponds to $2^{\alpha-2} y_c$, and the $y_c$ 
values on the horizontal axis indicate that we set $\lambda \le 0.3$.

The fixed-order results diverge as $y_c$ approaches zero, but the resummed results are suppressed for small $y_c$ and converge to zero.
The theoretical uncertainty at NLL order is not significantly reduced compared to the NLO fixed-order results. We expect that the uncertainty
will be reduced at higher orders. In addition to the suppression of the resummed result for small $y_c$, the fixed-order results tend to be more
negative as $\alpha$ decreases, while the jet cross sections have meaning only when they are positive. As Figure~\ref{csec} shows, the NLO result
and the NLL result are consistent except for small values of $y_c$ for $\alpha =1$. As $\alpha$ decreases, the fixed-order result is not reliable. 
Only after the resummation is performed, the jet cross sections become positive, including the behavior for small $y_c$. It turns out that 
the cone-type algorithm, which is equivalent to the inclusive $k_T$ algorithm at NLO, and the Sterman-Weinberg algorithm show similar shapes
for the phase spaces \cite{Chay:2015ila}, therefore we expect that the dijet cross section should be resummed in the cone-type algorithm to yield a 
meaningful answer.

Since the dijet cross section is IR finite in spite of the fact that both the jet and soft functions are IR divergent, one might expect that
the resummed result for the $k_T$ algorithm ($\alpha=2$) can be obtained by taking the limit $\alpha \rightarrow 2$ from the result with
$0<\alpha<2$. Indeed we can obtain the resummed result and verify that the cross section approaches zero for small $y_c$. However, the theoretical 
uncertainties blow up in this limit due to the large uncertainties in the jet and soft functions as $\alpha \rightarrow 2$, and there is no 
predictive power in the resummed result for $\alpha=2$.

\section{Conclusion\label{conc}}
The generalized jet algorithm enables us to see the rich structure of the jet algorithm from the viewpoint of the factorization. 
It includes familiar jet algorithms with specific values of $\alpha$. The advantage of the generalized jet algorithm is to see the divergence 
structure of the jet and soft functions in a unified way.  We can see systematically how the structure of the divergence changes as $\alpha$ 
varies.

The structure of the divergence can be inferred from the observation of the phase spaces in $(l_-, l_+)$ space. When it is 
ambiguous for $1<\alpha < 2$, the phase space in $(l_-, \mathbf{l}_{\perp}^2)$ space is helpful in identifying the divergence. In explicit
computation, we have verified that the expectation from the phase space is correct.

To summarize, the jet algorithm for $\alpha <2$ renders the jet and soft functions IR finite, thus guarantees the factorization of the dijet cross 
section. Among the known jet algorithms, this includes the JADE algorithm (corresponding to $\alpha=1$), the Cambridge/Aachen algorithm
($\alpha=0$) and the Sterman-Weinberg and the cone-type algorithm ($\alpha<0$) and the anti-$k_T$ algorithm ($\alpha=-2$). On the other hand,
the factorization breaks down with the jet algorithms with $\alpha \ge 2$ because the jet and soft functions contain infrared divergence. 
Though the IR divergence is cancelled in the dijet cross section and gives a finite result, it breaks the factorization and the $k_T$ algorithm 
belongs to this category.  

The main point of the paper is that the structure of divergence in the collinear and soft parts depends on how the jets are defined. The important
criterion is that each factorized part should be IR finite in order to respect the factorization. And according to our analysis of the
generalized jet algorithm, only the jet algorithm with $\alpha<2$ factorizes the dijet cross section in a strict sense. 

Note that we have considered the so-called exclusive generalized $k_T$ algorithms \cite{Catani:1991hj}, which is based on
the $e^+ e^-$ annihilation. The SCET version is presented in Ref.~\cite{Cheung:2009sg}. The inclusive recombination algorithm \cite{Cacciari:2008gp} 
is more suitable in hadronic collisions. But it turns out that the inclusive jet algorithm reduces to the cone-type algorithm \cite{Ellis:2010rwa}
at next-to-leading order in which there are three final-state particles. In contrast to $e^+ e^-$ annihilation, the partonic center of energy is boosted along the beam direction 
compared to the hadron-hadron center of energy in hadronic collisions. 
Therefore the physical observables in hadron-hadron scattering, as in LHC, should be described in terms of the boost-invariant quantities such as the 
transverse momentum, the rapidity and the azimuthal angle.  It will be interesting to probe the jet properties arising from the jet algorithms
in other types of scattering.
And the analysis of the divergence structure in various jet algorithms, which we have probed here, 
can be extended to deep inelastic scattering and hadron-hadron collisions with the boost invariance along the beam direction 
 to see if the structure of the divergence is sustained or if the kinematics of the scattering changes it using the exclusive
or inclusive generalized $k_T$ algorithm. 

\acknowledgments

J. Chay and I. Kim are supported by Basic Science Research Program through the National Research Foundation of Korea (NRF) funded by the Ministry of Education(Grant No. NRF-2014R1A1A2058142).   C.~Kim is supported by Basic Science Research 
Program through NRF 
funded by the Ministry of Science, ICT and Future Planning with  Grants No. NRF-2012R1A1A1003015,  No. NRF-2014R1A2A1A11052687.

\appendix
\section*{Ingredients in obtaining the NLL results}
The QCD $\beta$ function and the cusp anomlous dimension  in the $\overline{\mathrm{MS}}$ scheme are given as
\begin{eqnarray}
\beta(\alpha_s) &=& \frac{d}{d\ln \mu} \alpha_s = -2 \alpha_s \Bigl[ \beta_0 \frac{\alpha_s}{4\pi} 
+ \beta_1 \Bigl(\frac{\alpha_s}{4\pi}\Bigr)^2 +\cdots \Bigr], \nonumber \\
\Gamma_{\mathrm{cusp}} (\alpha_s) &=& \Gamma_0 \frac{\alpha_s}{4\pi} +\Gamma_1 \Bigl(\frac{\alpha_s}{4\pi}\Bigr)^2+\cdots, 
\end{eqnarray}
where the expansion coefficients for the QCD $\beta$ function to two-loop order are 
\begin{equation}
\beta_0 = \frac{11}{3}C_A -\frac{4}{3} T_F n_f, \ \beta_1 = \frac{34}{3}C_A^2 -\frac{20}{3} C_A T_F n_f -4C_F T_F n_f.
\end{equation}
And the cusp anomalous dimension to two loop order are given as \cite{Korchemsky:1987wg,Korchemskaya:1992je}
\begin{equation}
\Gamma_0 = 8C_F,  \ \ \Gamma_1 = 8C_F\Bigl[\Bigl( \frac{67}{9} -\frac{\pi^2}{3}\Bigr) C_A -\frac{20}{9}T_F n_f\Bigr].
\end{equation}

The coefficients of the anomalous dimensions in the hard function are given by
\begin{equation}
a_H^0= \Gamma_0, \ a_H^1 = \Gamma^1, \ b_H^0 = -12 C_F
\end{equation}
For the jet and the soft functions with $0<\alpha<2$, the coefficients are given a
\begin{eqnarray}
&& a_J^0 = \frac{1}{\alpha-2}\Gamma_0, \ a_J^1 = \frac{1}{\alpha-2}\Gamma_1, \ b_J^0 = 6C_F, \nonumber \\
&& a_S^0 = -\frac{\alpha}{\alpha-2} \Gamma_0, a_S^1 = -\frac{\alpha}{\alpha-2} \Gamma_1, b_S^0 = 0, 
\end{eqnarray}
and with $\alpha \le 0$, they are given as
\begin{eqnarray}
&& a_J^0 = -\frac{1}{2}\Gamma_0, \ a_J^1 = -\frac{1}{2}\Gamma_1, \ b_J^0 = 6C_F, \nonumber \\
&& a_S^0 = 0, \ a_S^1 = 0, \ b_S^0 = 8C_F \ln (2^{\alpha-2} y_c). 
\end{eqnarray}

At NLL order, the quantities in Eq.~(\ref{rgf}) are written as
\begin{eqnarray}
S_i (\mu_i,\mu) &=& \frac{a_i^0}{4\beta_0^2} \Bigl[ \frac{4\pi}{\alpha_s (\mu_i)} \Bigl( 1-\frac{\alpha_s (\mu_i)}{\alpha_s(\mu)}
-\ln \frac{\alpha_s (\mu)}{\alpha_s (\mu_i)} \Bigr)  +\frac{\beta_1}{2\beta_0} \ln^2 \frac{\alpha_s (\mu)}{\alpha_s (\mu_i)} 
\nonumber \\
&&+\Bigl( \frac{a_i^1}{a_i^0}-\frac{\beta_1}{\beta_0} \Bigr) 
\Bigl( 1- \frac{\alpha_s (\mu_i)}{\alpha_s (\mu)}  +\ln \frac{\alpha_s (\mu)}{\alpha_s (\mu_i)} \Bigr)\Bigr], \nonumber \\
A_i (\mu_i,\mu) &=& \frac{a_i^0}{2\beta_0} \Bigl[ \ln \frac{\alpha_s (\mu)}{\alpha_s (\mu_i)} + \Bigl( \frac{a_i^1}{a_i^0} 
-\frac{\beta_1}{\beta_0} \Bigr) \frac{\alpha_s (\mu) -\alpha_s (\mu_i)}{4\pi} \Bigr],
\end{eqnarray}
and $B_i (\mu_i, \mu)$ is obtained from $A_i (\mu_i,\mu)$ when $a_i^j$ are replaced by $b_i^j$.

\end{document}